\documentclass[12pt,twoside,a4paper]{memoir}

\usepackage[utf8]{inputenc}   %cosas del codigo
\usepackage[T1]{fontenc}      %letras acentuadas
\usepackage{amsmath}          %insertar ecuaciones alineadas
\usepackage{amsthm}           %algo de teoremas
\usepackage{amssymb}          %escribir simbolos matematicos
\usepackage{mathtools}        %para mejorar los documentos matematicos
\usepackage{multirow}         %combinar filas de tablas en una sola celda
\usepackage{tabularx}         %definir tablas de ancho fijo
\usepackage [all]{xy}         %para graficos y matrices
\usepackage{graphicx}         %incluir imagenes
\usepackage{float}            %para figuras y tablas (marcos, posiciones...)
\usepackage{rotating}         %para rotar graficos y figuras
\usepackage{booktabs}         %mejorar tablas (comandos extra)
\usepackage[symbol]{footmisc}  
\usepackage{lmodern}
\usepackage{parskip}
\usepackage{tikz}
\usepackage{amsmath,amssymb,amstext,amsfonts}
\usepackage{eucal}
\usepackage{wrapfig}   %Figuras no alineadas con el texto
\usepackage{lipsum}
\usepackage{caption}    %añadir subfiguras
\usepackage{subcaption} %crear grupos de figuras o imagenes relaciomadas
\usepackage{hyperref}     %enlaces y documentos hipertextuales

\hypersetup{colorlinks=true}  %usar colores
\usepackage{color}     %colorines
\usepackage{endnotes}  %poner notas a pie de pagina
\usepackage{setspace}  %Ajustar interlineado
\usepackage{array}     %cosas de tablas
\usepackage{textcomp}  %Simbolos varios
\usepackage{afterpage} %Para poner una pagina en blanco
\usepackage{emptypage} %para no numerar una pagina
\usepackage{mathrsfs}  %mas simbolos
\usepackage{enumerate} %para enumeraciones
\usepackage[scaled]{beramono}  %fuentes de letras
\usepackage{listingsutf8} %bibliografia
\usepackage{verbatim}  %para añadir comentarios, basicamente como el %
\usepackage{tikz}
\usepackage{mathrsfs}

\usepackage{amsmath}
\usepackage{amssymb}
\usepackage{amsfonts}
\usepackage{latexsym}
\usepackage{cancel}
\usepackage{rawfonts}
\usepackage{pictexwd}

\definecolor{anti-flashwhite}{rgb}{0.95, 0.95, 0.96}
\definecolor{bananamania}{rgb}{0.98, 0.91, 0.71}
\definecolor{caribbeangreen}{rgb}{0.0, 0.8, 0.6}
\definecolor{almond}{rgb}{0.94, 0.87, 0.8}
\definecolor{antiquewhite}{rgb}{0.98, 0.92, 0.84}
\definecolor{babyblue}{rgb}{0.19, 0.55, 0.91}

\newtheorem{teo}{Theorem}[chapter]
\newtheorem{lema}{Lemma}[chapter]
\newtheorem{prop}{Proposition}[chapter]

\newtheorem{defi}{Definition}[chapter]
\newtheorem{obs}{Remark}[chapter]
\newtheorem{ej}{Example}[chapter]

\usepackage{stmaryrd}
\usepackage{libertine}
\usepackage[libertine,cmintegrals,cmbraces,vvarbb]{newtxmath}
\usepackage[scaled=0.6]{zi4}
\AtBeginDocument{%
  \let\mathbb\relax
  \DeclareMathAlphabet{\mathbb}{U}{msb}{m}{n}%
}

\newcommand*{\R}{\mathbb{R}}

\newcommand*{\Lie}{\mathscr{L}}
\newcommand*{\RR}{\mathscr{R}}
\newcommand*{\HH}{\mathscr{H}}

\usepackage[left=3cm,right=3cm,top=3cm,bottom=3cm]{geometry}

\begin{document}

\thispagestyle{empty}
\vspace*{0.01cm}
\begin{center}
{\Large \textbf{UNIVERSIDAD COMPLUTENSE DE MADRID}} 

\vspace{0.5cm}
{\large \textbf{FACULTAD DE CIENCIAS MATEMÁTICAS}}

\vspace{1.2cm}
\begin{figure}[h]
\centering
\includegraphics{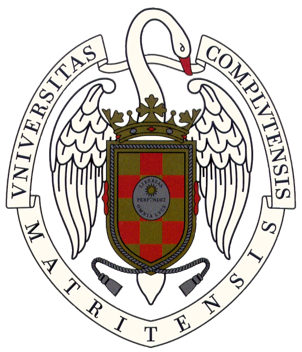}
\end{figure} 

\vspace{0.5cm}
{\large {TRABAJO FIN DE MÁSTER}}

\vspace{1.5cm}
{\huge MOMENTUM MAPPING AND}

\vspace{0.2cm} 
{\huge REDUCTION IN CONTACT} 

\vspace{0.2cm} 
{\huge HAMILTONIAN SYSTEMS}

\vspace{1.5cm}
{\large MÁSTER EN MATEMÁTICAS AVANZADAS}

\vspace{0.8cm}
{\Large Juan Manso García-Mauriño}

\vspace{0.5cm}
{\large Director: Manuel de León Rodríguez (ICMAT)}

\vspace{0.3cm}
{\large Tutor: Marco Castrillón López (UCM)}

\vspace{0.8cm}
\begin{large}
Curso Académico 2021-2022
\end{large}
\end{center} 
\chapter*{Abstract}
\thispagestyle{empty}

Due to the emergence of symplectic geometry, the geometric treatment of mechanics underwent a great development during the last century. In this scenario the pressence of symmetries in Hamiltonian systems leads naturally to the existence of conserved quantities. This integrals of motions are described by the well-known momentum mapping. Furthermore, the equations of motion can be simplified by a process known as reduction, if the system is invariant under the action of a certain group. This process can also be considered in the framework of contact geometry, a much more recent field of study. This kind of geometry has proven to be valuable in areas as different as thermodynamics, control theory, or neurogeometry. Our knowledge about contact geometry is much smaller than that of symplectic, and so a process known as symplectificacion is extremely useful, since it allows to study contact problems in the symplectic frame.

The aim of the text is to study the commutativity relations between the processes of reduction and symplectification. To do so, we first introduce the basis of symplectic and contact geometry and we show how to perform the reduction via the momentum map in both scenarios. The well-known coisotropic reduction theorem will be crucial in the description. Finally, the symplectification process is analyzed in detail, and its relation with symplectic and contact reduction is studied.

\chapter*{Resumen}
\thispagestyle{empty}

El tratamiento geométrico de la mecánica experimentó un gran desarrollo durante el último siglo, gracias al nacimiento de la geometría simpléctica. Dentro de este marco es sencillo mostrar que la presencia de simetrías en un cierto sistema Hamiltoniano conduce a la existencia de cantidades conservadas. Estas integrales del movimiento quedan codificadas en la conocida como aplicación momento. Además, a través de un proceso denominado reducción, es posible simplificar las ecuaciones del movimiento de un sistema que posee simetrías. Este mismo proceso puede llevarse a cabo en los llamados sistemas de contacto. Durante los últimos años se ha producido un gran interés en las aplicaciones de esta clase de geometría, que ha demostrado ser útil en ámbitos tan distintos como la termodinámica, la teoría de control, o la neurogeometría. El escaso desarrollo de la geometría de contacto en comparación con la simpléctica convierte en esencial un proceso conocido como simplectificación, que permite transformar problemas de contacto en simplécticos. 

El objetivo del texto es estudiar las relaciones de conmutatividad entre los procesos de reducción y simplectificación. Para ello, primero se realiza una introducción a las geometrías simpléctica y de contacto, y se expone la forma en que la reducción es llevada a cabo en cada caso. Para ello será especialmente relevante el conocido como teorema de reducción coisotrópica. Por último se analiza en detalle el proceso de simplectificación y se estudia su relación con las reducciones simpléctica y de contacto.
  
\newpage 
\thispagestyle{empty}
\ 

\newpage

\tableofcontents*
\thispagestyle{empty}

\chapter*{Introduction}
\addcontentsline{toc}{chapter}{Introduction} \setcounter{page}{1}
The systematic mathematical study of mechanics goes back to Newton. Later on, the development of the Lagrangian and Hamiltonian formalisms represented a great advance in the discipline. But it is not until the early twentieth century when geometry comes into play, leading to the birth of what we know as geometric mechanics. This new approach led to a spectacular advance in mechanics, especially in the second half of the 20th century, where symplectic geometry was thoroughly study. In this framework, a phase space is described by a manifold endowed with a symplectic form, and the dynamics of the system is given by the integral curves of a certain vector field.

The relationship between symmetries and constants of motion is crucial in the analysis of physical systems. Indeed, the presence of symmetries reduces the degrees of freedom of the system, leading to a simplification of its equation of motion. Emmy Noether works were decisive in the study of the mentioned relation, but the modern formulation of the problem is based on what is known as momentum mappings. Linear and angular momentum are some paradigmatic examples. Using this momentum maps, the simplification of systems with symmetries under the action of a certain Lie group can be made by a systematic procedure called reduction \cite{Marsden:1974dsb}.  

Many physical systems, such as dissipative systems, can not be described by the symplectic formalism, and other geometric structures are needed. In this context, there has been a growing interest in contact geometry during the last few years. Apart from be the natural scenario to describe dissipative systems, contact geometry is applicable in many other areas, such as thermodynamics \cite{bravetti2019contact, 10.1007/978-3-030-77957-3_13}, cosmology \cite{sloan2021new}, control theory \cite{ohsawa2015contact}, or neurogeometry \cite{petitot2017elements}. See reference \cite{bravetti2017contact} for a recent review of several of these topics. The relation between symmetries and dissipated quantities has also been studied in this context. Moreover, appropiate momentum maps allow us to reduce the dynamics of contact Hamiltonian systems \cite{CHS}.

There exists a mechanism to treat contact systems as symplectic ones. This procedure is known as symplectification \cite{ibanez1997co}. It is a very powerful tool, because our knowledge of symplectic geometry is much greater than that of contact geometry. Since the reduction of a contact system is another contact system, a natural question arises: does the symplectification of the reduced contact space coincide with the symplectic reduction of the symplectification of the original contact space?

The aim of the text is to give an answer to that question. To do so, we first introduce the basis of symplectic and contact geometry and we show how to perform the reduction via the momentum map in both scenarios. Then we study some properties of the symplectification process and finally we clarify the relation between simplictification and reduction. 

Chapter \ref{chapter1} is dedicated to symplectic geometry. In section \ref{sec1.1} we first recall the basis of the theory. Then we introduce Hamiltonian systems, and finally we consider some relevant types of submanifolds of symplectic manifolds, namely, isotropic, coisotropic, and Lagrangian submanifolds. Section \ref{sec1.2} is a previous step for proving the symplectic reduction via the momentum map. There, we prove the important coisotropic reduction theorem due to Weinstein \cite{Marsden:1974dsb}, and we show that Lagrangian submanifolds are projected onto Lagrangian submanifolds of the reduced space in certain circunstances. Section \ref{sec1.3} is devoted to symplectic reduction via the momentum map. We first define momentum mappings and give some examples, and then we use the coisotropic reduction theorem to perform the reduction.

Chapter \ref{chapter2} is dedicated to contact geometry. Its structure is similiar to that of chapter \ref{chapter1}. In section \ref{sec2.1} we first introduce contact forms and contact Hamiltonian systems. Then Jacobi manifolds are introduced, and symplectic and contact manifolds are presented as examples of this more general setting. Sections \ref{sec2.2} and \ref{sec2.3} are the contact counterparts of sections \ref{sec1.2} and \ref{sec1.3}.

Chapter \ref{chapter3} contains some new result. In section \ref{sec3.1} we introduce the symplectification process and show that coisotropic submanifolds are transformed into coisotropic submanifolds. In section \ref{sec3.2} we state the commutativity of symplectification and coisotropic reduction. Finally, this is used to prove the analogue result in the context of momentum mappigs in section \ref{sec3.3}.
\chapter{Reduction in symplectic geometry}\label{chapter1}
\section{Symplectic geometry}
Symplectic geometry is the natural arena for Hamiltonian mechanics. Some reference textbooks are \cite{abraham2008foundations,arnold1997mathematical,de1989methods,libermann1987symplectic}. In this section we introduce the basis of the theory and we endow physical systems with a symplectic structure.

Consider a manifold $M$ and let $\omega$ be a $2$-form on $M$. For each $x\in M$ we can consider a map from the tangent space $T_xM$ to its dual $T^*_xM$, namely
\begin{align*}
\flat_x:T_xM&\rightarrow T_x^*M,\\ u&\mapsto i_u\omega.
\end{align*}
The dimension of the image of $\flat_x$ is called the \textbf{rank} of $\omega$ at $x$. It can be shown that the rank of every $2$-form is an even number. We say that a $2$-form $\omega$ is nondegenerate if, for all $x\in M$, the map $\flat_x$ is an isomorphism, or equivalently,
$$\text{rank}\,\omega(x)=\dim (\text{im}\, \flat_x)=\dim(T_x^*M)=\dim M.$$
In this sense, nondegenerate forms can be defined only on even dimensional manifolds. 

\begin{defi}(Symplectic manifold). A symplectic manifold is a pair $(M,\omega)$, where $M$ is a $2n$-dimensional manifold and $\omega$ is a symplectic form, that is, a closed nondegenerate $2$-form on $M$.
\end{defi}

Every symplectic manifold induces a natural vector bundle isomorphism. In fact, if $(M,\omega)$ is a symplectic manifold, we have
\begin{align*}
\flat:TM&\rightarrow T^*M,\\ u&\mapsto i_u\omega.
\end{align*}
The inverse is denoted by $\sharp=\flat^{-1}$. We can also see $\flat$ as an isomorphism between vector fields and $1$-forms on $M$.

\begin{defi}(Symplectic transformation). Let $(M,\omega),(N,\alpha)$ be symplectic manifolds of the same dimension. A differentiable mapping $F:M\rightarrow N$ is said to be a symplectic transformation if it preserves the symplectic forms, that is,
$$F^*\alpha=\omega.$$ 
Moreover, if $F$ is a global diffeomorphism, we called it symplectomorphism. In particular, when $M=N$ and $F:M\rightarrow M$ preserves the symplectic form, $F$ is said to be a canonical transfomation.
\end{defi}

Next we present the paradigmatic example of symplectic manifold, which is that of a phase space, or what is the same, the cotangent bundle of a given configuration manifold. Let $M$ be a manifold, $T^*M$ its cotangent bundle, and $\pi_M:T^*M\rightarrow M$ the canonical projection. We define the $1$-form $\lambda_M$ on $T^*M$ by,
$$\lambda_M(\alpha)(u)=\alpha(T_{\alpha}\pi_M(u)),$$
for $\alpha\in T^*_xM$, $u\in T_\alpha(T^*M)$. $\lambda_M$ is called the Liouville form on $T^*M$.
Locally, if $(q^i)$ are coordinates on $M$ and $(q^i,p_i)$ are the induced coordinates in $T^*M$, we have
\begin{align*}
\lambda_M(q^j,p_j)(\partial/\partial q^i)&=(p_jdq^j)(d_{\alpha}\pi_M(\partial/\partial q^i))=(p_jdq^j)(\partial/\partial q^i)=p_i,\\
\lambda_M(q^j,p_j)(\partial/\partial p_i)&=(p_jdq^j)(d_{\alpha}\pi_M(\partial/\partial p_i))=(p_jdq^j)(0)=0.
\end{align*}
Thus,
$$\lambda_M=p_idq^i.$$
The Liouville form is the unique $1$-form on $M$ such that
$$\alpha^*\lambda_M=\alpha,$$
for all $1$-form $\alpha$ on $M$. 

It can be easily shown that the $2$-form,
$$\omega_M=-d\lambda_M,$$
defined locally by,
\begin{equation}\label{formadarboux}
\omega_M=dq^i\wedge dp_i,
\end{equation}
is a symplectic form in $T^*M$, which is known as the canonical symplectic form on $T^*M$.

Darboux theorem states that the above local expression $\eqref{formadarboux}$ is not exclusive of the canonical symplectic form on $T^*M$.

\begin{teo}(Darboux theorem). Let $(M,\omega)$ be a $2n$-dimensional symplectic manifold. There exists a neighbourhood of each point with coordinates $(q^i,p_i)$ such that the symplectic form can be written as
$$\omega=dq^i\wedge dp_i.$$
\end{teo}

\subsection{Hamiltonian systems}
The symplectic structure just defined allows us to introduce dynamics in symplectic manifolds, when a Hamiltonian function is given.
\begin{defi}(Hamiltonian system). Let $H:M\rightarrow \R$ be a smooth map on a symplectic manifold $(M,\omega)$. We define the associated Hamiltonian vector field $X_H$ by,
\begin{equation}\label{XHsym}
X_H=\sharp(dH),
\end{equation}
or equivalenty,
\begin{equation}\label{dHsym}
i_{X_H}\omega=dH.
\end{equation}
The triple $(M,\omega,H)$ is said to be a Hamiltonian system, and $H$ is usually called energy (or Hamiltonian energy).
\end{defi}

Now we study the trajectories of a Hamiltonian system $(M,\omega,H)$, that is, the integral curves of $X_H$. Let $(q^i,p_i)$ be Darboux coordinates in $M$ and let $\sigma :I=(-\epsilon,\epsilon)\rightarrow M$ be an integral curve of $X_H$, that is, 
\begin{equation}\label{curvaintegral}
X_H(\sigma(t))=\dot{\sigma}(t),\quad t\in I.
\end{equation}
In local coordinates we write $\sigma(t)=(q^i(t),p_i(t))$, so that
\begin{equation}\label{sigma}
\dot{\sigma}(t)=\dfrac{dq^i}{dt}\dfrac{\partial}{\partial q^i}+\dfrac{dp_i}{dt}\dfrac{\partial}{\partial p_i}.
\end{equation}
On the other hand, $dH$ can be written locally as,
$$dH=\dfrac {\partial H}{\partial q^i}dq^i+\dfrac{\partial H}{\partial p_i}dp_i,$$
and by definition \eqref{XHsym},
$$X_H=\sharp(dH)=\dfrac {\partial H}{\partial q^i}\sharp(dq^i)+\dfrac{\partial H}{\partial p_i}\sharp(dp_i).$$
But in Darboux coordinates $\omega=dq^i\wedge dp_i$. A simple computation shows that
$$\flat\left(\dfrac{\partial}{\partial q^i}\right)=dp_i, \quad\, \flat\left(\dfrac{\partial}{\partial p_i}\right)=dq^i, \quad\, \sharp(dq^i)=-\dfrac{\partial}{\partial p_i}, \quad\, \sharp(dp_i)=\dfrac{\partial}{\partial q^i}.$$
Consequently,
\begin{equation}\label{XH}
X_H=\dfrac {\partial H}{\partial p_i}\dfrac{\partial}{\partial q^i}-\dfrac{\partial H}{\partial q^i}\dfrac{\partial}{\partial p_i}.
\end{equation} 
Finally, integral curves of $X_H$ are given by
\begin{align*}
\dfrac{dq^i}{dt}&=\dfrac {\partial H}{\partial p_i},\\ \dfrac{dp_i}{dt}&=-\dfrac {\partial H}{\partial q^i},
\end{align*}
which are the well-known Hamilton equations. This is a crutial result, since it justifies the use of symplectic geometry in the study of mechanics.

\begin{ej}(Particle in a potential)
Let $M=\mathbb{R}^3$ be the configuration space of a particle. The phase space (positions and momentum) is then given by the cotagent bundle $T^*M$ of $M$. Consider the sympletic manifold $(T^*M,\omega_M)$ and let $H:T^*M\rightarrow \mathbb{R}$ be the function,
\begin{equation}\label{hamiltonianopotencial}
H(q^1,q^2,q^3,p_1,p_2,p_3)=\dfrac{1}{2m}\sum_{i=1}^3p_i^2+U(q^1,q^2,q^3),
\end{equation}
where $U$ is a differentiable map describing a potential in $\mathbb{R}^3$. Hamilton equations are then given by
\begin{align*}
\dot{q}^i&=\dfrac {p_i}{m},\\ \dot{p}_i&=-\dfrac {\partial U}{\partial q^i}.
\end{align*}
Thus we have,
$$m \ddot{q}^i=-\dfrac {\partial U}{\partial q^i},$$
that is, Newton's second law for a particle of mass $m$ moving in a potential.

If, for instance, $U$ is the armonic potential
$$U(q_1,q_2,q_3)=\dfrac{1}{2}k\textbf{q}^2,$$
we get
$$m \ddot{q}^i=-kq^i,$$
which is the well-known Hooke's low. $\hfill\diamondsuit$
\end{ej}

The energy function $H$ is preserved during the evolution of the system. Indeed,
$$\mathscr{L}_{X_H}H=X_H(H)=dH(X_H)=i_{X_H}\omega(X_H)=\omega(X_H,X_H)=0,$$
that is, $H$ is constant along the integral curves of $X_H$. 

More generally, if $X$ is a vector field on $M$ and $f:M\rightarrow \R$  is a map such that $X(f)=0$ we say that $f$ is a first integral of $X$. Similarly, a $1$-form $\alpha$ on $M$ is called a first integral if $\alpha(X)=0$. In the particular case when $f$ is a first integral of $X_H$, we called it a constant of the motion. 

\begin{ej}(Conservation of angular momentum). Let $H$ be the Hamiltonian in the above example, \eqref{hamiltonianopotencial}, and suppose that $U$ depends only on the distance from the origin, that is, $U=U(r)$, where $r=\|(q^1,q^2,q^3)\|$. By \eqref{XH}, the Hamiltonian vector field $X_H$ is given by
$$X_H=\dfrac{p_i}{m}\dfrac{\partial}{\partial q^i}-\dfrac{U'(r)}{r}q^i\dfrac{\partial}{\partial p_i}.$$
Now we consider the function $L_3=q^1p_2-q^2p_1$. Clearly,
$$X_H(L_3)=\dfrac{p_1}{m}p_2-\dfrac{p_2}{m}p_1-\dfrac{U'(r)}{r}q^1(-q^2)-\dfrac{U'(r)}{r}q^2q^1=0,$$
thus $L_3$ is a constant of the motion. Similarly, $L_1=q^2p_3-q^3p_2$ and $L_2=q^3p_1-q^1p_3$ are also constants of the motion. We conclude that the angular momentum $\textbf{L}=\textbf{q}\times \textbf{p}$ is conserved, as it was expected for a rotationally invariant Hamiltonian.  $\hfill\diamondsuit$
\end{ej}

\subsection{Submanifolds of a symplectic manifold}
We next consider some relevant types of submanifolds of symplectic manifolds. To do so, we first consider a notion of orthogonality in each tangent space.
\begin{defi}\label{orthocomplement} Let $(M,\omega)$ be a symplectic manifold, $x\in M$, and $\Delta_x$ a linear subspace of $T_xM$. Then the subspace,
$$\Delta_x^{\perp}=\sharp(\Delta_x^\circ),$$
is called the orthocomplement of $\Delta_x$ in $T_xM$ (with respect to $\omega$). Here $\Delta_x^\circ=\lbrace\alpha\in T_xM^*\,|\,\alpha(\Delta_x)=0\rbrace$ is the annihilator of $\Delta_x$. Note that we can write,
$$\Delta_x^{\perp}=\lbrace v\in T_xM\,|\, \omega(u,v)=0\; \text{for all}\; v\in \Delta_x\rbrace.$$
\end{defi}
One has,
\begin{equation}\label{dimensionesorto}
dim T_xM=\dim \Delta_x+\dim\Delta_x^\perp.
\end{equation}
Indeed, since $\sharp$ is an isomorphism,
$$\dim\Delta_x^\perp=\dim \sharp(\Delta_x^\circ)=\dim \Delta_x^\circ=\dim T_xM-\dim \Delta_x.$$
Using both this and the properties of anihilators, the following identities can be easily obained
\begin{align}
(\Delta_x^\perp)^\perp&=\Delta_x, \\
\label{dedemorgan}(\Delta_x\cap\Pi_x)^\perp&=\Delta_x^\perp+\Pi_x^\perp,\\
(\Delta_x+\Pi_x)^\perp&=\Delta_x^\perp\cap\Pi_x^\perp,
\end{align}
where $\Delta_x,\Pi_x$ are linear subspaces of $T_xM$.

Definition \ref{orthocomplement} can be extented to distributions $\Delta\subset TM$ and submanifolds $N\subset M$ by taking the complement pointwise in each tangent space. The resulting distribution, noted by $TN^\perp$, is usually called the characteristic distribution of $N$.

\begin{defi} A submanifold $N$ of a symplectic manifold $(M,\omega)$ is called,
\begin{itemize}
\item[i)] isotropic if $T_xN\subset T_xN^{\perp}$,
\item[ii)] coisotropic if $T_xN^{\perp}\subset T_xN$,
\item[iii)] Lagrangian if $T_xN$ is a maximal isotropic subspace of $T_xM$,
\item[iv)] symplectic if $T_xN\cap T_xN^\perp=0$,
\end{itemize}
for all $x\in N $.
\end{defi}

The following two propositions are characterizations of the concept of Lagrangian submanifold.
\begin{prop} Let $(M,\omega)$ be a $2n$-dimensional symplectic manifold. A submanifold $N\subset M$ is Lagrangian if and only if $T_xN=T_xN^\perp$ for all $x\in N$. 
\end{prop}  
\begin{proof}
Let $x\in N$ and suppose by way of contradiction that $T_xN\subset T_xN^\perp$ but $T_xN^\perp\backslash T_xN\neq 0$. We can choose $u\in T_xN^\perp\backslash T_xN$ and then we have,
$$\omega(v+au,w+bu)=0,$$
for all $v,w\in T_xN$ and $a,b\in \R$. This implies that $T_xN+\langle u\rangle$ is isotropic, contradicting the assumption that $T_xN$ is maximal. 

To prove the converse, note that \eqref{dimensionesorto} and $T_xN=T_xN^\perp$ together imply $\dim T_xN=n$. Also by \eqref{dimensionesorto}, $n$ is the maximum dimension possible for a isotropic submanifold, thus $T_xN$ is maximal and $N$ is Lagrangian.
\end{proof}

\begin{prop}\label{nsymplectic} Let $(M,\omega)$ be a $2n$-dimensional symplectic manifold. A submanifold $N\subset M$ is Lagrangian if and only if it is isotropic and $\dim N=n$. 
\end{prop}
\begin{proof}
Let $x\in N$. If $N$ is Lagrangian then $T_xN=T_xN^\perp$, so using \eqref{dimensionesorto} we are done. Conversely, if $\dim T_xN=n$, then also $\dim T_xN^\perp=n$, and so $T_xN=T_xN^\perp$ because $N$ is isotropic.
\end{proof}\label{sec1.1}
\section{Symplectic coisotropic reduction}
In this section we study the well-known coisotropic reduction theorem due to Marsden and Weinstein \cite{Marsden:1974dsb}. However, our scheme is essentially based on \cite{CHS}.

\begin{teo}\label{symcoired}(Coisotropic reduction). Let $i:N\hookrightarrow M$ be a coisotropic submanifold. Then, $TN^\perp$ is an involutive distribution on $N$.

Then, Frobenius theorem guarantees that the distribution $TN^\perp$ is completely integrable, thus, it defines a foliation on $N$. Assume that the obtained leaf space, denoted by $\tilde{N}=N/TN^\perp$, is a manifold. Let $\pi:N\rightarrow \tilde{N}$ be the projection. Then, there exists a unique $2$-form $\tilde{\omega}\in \Omega^2(\tilde{N})$ such that,
\begin{equation}\label{symred1}
\pi^*\tilde{\omega}=i^*\omega \coloneqq\omega_0.
\end{equation}
Furthermore, $\tilde{\omega}$ is symplectic and $(\tilde{N},\tilde{\omega})$ is a symplectic manifold.
\end{teo}

\begin{proof}First, we will show that $TN^\perp$ is an involutive distribution. Let $X_1,X_2\in TN^\perp$. We have to prove that $\left[X_1,X_2\right]$ takes values in $TN^\perp$ as well. However, $\omega(X_i,Y)=0$ for all vector fields $Y\in \chi(N)$. Hence, by proposition 1.12.3 \cite{de1989methods}
\begin{align*}
0&=d\omega(X_1,X_2,Y)=X_1(\omega(X_2,Y))-X_2(\omega(X_1,Y))+Y(\omega(X_1,X_2))
\\ &-\omega(\left[X_1,X_2\right],Y)+\omega(\left[X_1,Y\right],X_2)-\omega(\left[X_2,Y\right],X_1)=-\omega(\left[X_1,X_2\right],Y).
\end{align*}
We conclude that $[X_1,X_2]\in TN^\perp$.
\\

Now, we will prove that there is a unique $2$-form verifying \eqref{symred1}. For the uniqueness, let $p\in \tilde{N}$, $\tilde{u},\tilde{v}\in T_p\tilde{N}$,  and let $x\in \pi^{-1}(p)$ and $u,v\in T_xN$ (which exist because $\pi$ is a submersion) such that $T_x\pi(u)=\tilde{u}$, $T_x\pi(v)=\tilde{v}$. Suposse there is some $\tilde{\omega}$ such that $\pi^*\tilde{\omega}=\omega_0$.  Then,
\begin{equation}\label{tildeomega}
\tilde{\omega}(p)(\tilde{u},\tilde{v})=\tilde{\omega}(\pi(x))(T_x\pi(u),T_x\pi(v))=\pi^*\tilde{\omega}(x)(u,v)=\omega_0(x)(u,v).
\end{equation}
The right side does not depend on the $2$-form chosen, and then the $2$-form is unique in case it exists.
\\

For the existence, we will show that \eqref{tildeomega} defines the desired $2$-form. First, we must prove the independence on the chosen $x,u,v$. 

Fix $x\in \pi^{-1}(p)$. The independence in $T_xN$ follows from $\ker(T_x\pi)=(T_xN)^\perp$. Indeed, given $\bar{u},\bar{v}\in T_xN$ such that $T_x\pi(\bar{u})=\tilde{u}$, $T_x\pi(\bar{v})=\tilde{v}$, we have $u-\bar{u},v-\bar{v}\in \ker(T_x\pi)=(T_xN)^\perp$. Then,
$$\omega_0(x)(\bar{u},\bar{v})=\omega_0(x)(\bar{u},\bar{v})+\omega_0(x)(u-\bar{u},\bar{v})+\omega_0(x)(u,v-\bar{v})=\omega_0(x)(u,v).$$
Thus $\tilde\omega(p)$ does not depend on the chosen elements of the preimage of $T_x\pi$.

Now, we prove the independence on the point on the leaf chosen. Let $X$ be tangent to the leaf $\pi^{-1}(p)$, that is, $X\in TN^\perp$ so $i_X\omega_0=0$. Thus, by Cartan's formula,
$$\Lie_X\omega_0=di_x\omega_0+i_xd\omega_0=0.$$
We conclude that $\tilde{\omega}$ in \eqref{tildeomega}  is well defined.

Next we prove that $\tilde{\omega}$ is a $2$-form. Since $\pi$ is a submersion, there exists locally a smooth section $\mu:\tilde{N}\rightarrow N$ (that is, $\pi \circ \mu=id_{\tilde{N}}$). Take $\bar{\omega}=\mu^*\omega_0$. The independence in $\pi^{-1}(p)$ just checked gives $\pi^*\bar{\omega}=\omega_0$, and then \eqref{tildeomega} implies $\tilde{\omega}=\bar{\omega}$ wherever $\bar{\omega}$ is defined. This can be made in a neighborhood of each point, thus $\tilde{\omega}$ is a $2$-form. 

As $\tilde{\omega}$ satisfies \eqref{symred1} by definition, only last to prove that it is symplectic. Since $d$ and pullback operation commute, we have $\pi^*d\tilde{\omega}=i^*d\omega=0$, and then $d\tilde{\omega}=0$ because $\pi$ is a submersion. Hence, $\tilde{\omega}$ is closed.

Finally, we prove that $\tilde{\omega}$ is nondegenerate. Let $\tilde{u}\in T_{[p]}\tilde{N}$ such that,
$$\tilde{\omega}(p)(\tilde{u},\tilde{v})=0,$$
for all $\tilde{v}\in T_{p}\tilde{N}$. Let $x\in \pi^{-1}(p)$, $u\in T_xN$ such that $\tilde{u}=T_x\pi(u)$, and $v\in T_xN$, thus $T_x\pi(v)\in T_p\tilde{N}$. Hence, by \eqref{tildeomega},
$$\omega_0(x)(u,v)=\tilde{\omega}(p)(\tilde{u},T_x\pi(v))=0,$$
for all $v\in T_xN$. But this is only possible if $u\in (T_xN)^\perp=\ker(T_x\pi)$, and then $\tilde{u}=0$, so $\tilde{\omega}$ is nondegenerate. 

This proves that $\tilde{\omega}$ is symplectic and completes the prove. 
\end{proof}

As a consequence of the above theorem, we have that the projection of a Lagrangian submanifold is also a Lagrangian submanifold. 

\begin{prop}
Let $L$ be a Lagrangian submanifold, $N$ a coistropic submanifold, and $\pi:N\rightarrow \tilde{N}$ the asociated projection to the leaf space. Suppose that $L$ and $N$ have clean intersection, that is, $N\cap L$ is a submanifold and $T(N\cap L)=TN\cap TL$. Then the projection $\tilde{L}=\pi(L\cap N)\subset \tilde{N}$ is also a Lagrangian submanifold.
\end{prop}
\begin{proof}
Let $2n$ be the dimension of $M$, and $n+k$ the dimension of $N$. Since $L$ is Lagrangian, $dim\, L=n$, and using that $\omega$ is symplectic, we have $dim\, (TN^\perp)=n-k$. Hence,
$$dim \,\tilde{N}=dim\, N-dim (TN^\perp)=2k.$$
By proposition \ref{nsymplectic}, it suffices to prove that $\tilde{L}$ is isotropic and that $dim\,\tilde{L}=k$.
\\

First, we will prove that $\tilde{L}$ is isotropic. Let $[x]\in \tilde{L}$ and $\tilde{u}, \tilde{v}\in T_{[x]}\tilde{L}$. We will show that 
$$\tilde{\omega}([x])(\tilde{u},\tilde{v})=0,$$
so $\tilde{u}\in T_{[x]}\tilde{L}^\perp$ and then $T_{[x]}\tilde{L}\subset T_{[x]}\tilde{L}^\perp$.

Since $\pi$ is a submersion, there exist $u,v\in T_x(N\cap L)$ such that $\tilde{u}=T_x\pi(u), \tilde{v}=T_x\pi(v)$, and then,
$$\tilde{\omega}([x])(\tilde{u},\tilde{v})=\tilde{\omega}(\pi(x))(T_x\pi(u),T_x\pi(v))=\pi^* \tilde{\omega}(x)(u,v)=\omega (x)(u,v)=0,$$
since $\pi^*\tilde{\omega}=\omega$ and $L$ is Lagrangian.
\\

Now, to prove that $dim\,\tilde{L}=k$, note that,
\begin{equation}\label{abba1}
dim\,(N\cap L)+dim\,(N\cap L)^\perp=2n.
\end{equation}
By \eqref{dedemorgan}, we have
$$(N\cap L)^\perp=N^\perp+L^\perp=N^\perp+L,$$
and using Grassmann formula, 
$$dim\, (N^\perp+L)=dim\, N^\perp+dim\, L-dim\,(N\cap L^\perp)=2n-k-dim\,(N\cap L^\perp).$$

Substituting in \eqref{abba1}, we get,
$$dim\,(N\cap L)=2n-dim\,(N\cap L)^\perp=k+dim\,(N\cap L^\perp).$$
We conclude that, 
$$dim\,\tilde{L}=dim\,(N\cap L)-dim\,(N\cap L^\perp)=k.$$

\end{proof}

\label{sec1.2}
\section{Momentum map and symplectic reduction}
We have already study the relation between one-parameter groups of symmetries and constant of the motion in hamiltonian systems. But this is not the more general case, as we can deal with groups of symmetries which are not one-parameter groups. Then, we get a vector-valued conserved quantity called the momentum. For instance, translational and rotational invariance have linear and angular momentum as associated conserved quantities. The momentum map generalizes the situation to arbitrary groups of symmetries. 
This momentum map will also be useful to simplify the equations of motion. Indeed, it is well known that the number of degrees of fredoom of a system is reduced in presence of symmetry. In this context, we will prove a new reduction theorem, closely related with the coisotropic one. Some references for this section are \cite{abraham2008foundations,heckman2004geometry,Marsden1999,Marsden:1974dsb,ortega2013momentum}, and for basic Lie group theory one can see \cite{hall2003lie,warner1983lie}.

First of all, we should define the momentum map.

\begin{defi} \label{momentummap}\textbf{(Momentum map)}
Let $(M,\omega)$ be a connected sympletic manifold, $G$ a Lie group, and $\Phi:G\times M\rightarrow M$ a symplectic action of $G$ on $M$; that is, such that the map $\Phi_g:M\rightarrow M$ is a symplectic transformation for all $g\in G$. We say that a mapping
$$J:M\rightarrow \mathfrak{g}^*,$$ 
is a momentum map for the action $\Phi$ if, for all $\xi\in \mathfrak{g}$,
\begin{equation}\label{condmomento}
d\hat{J}(\xi)=i_{\xi_M}\omega,
\end{equation}
where $\hat{J}(\xi):M\rightarrow\mathbb{R}$ is given by $\hat{J}(\xi)(x)=J(x)\xi$, and $\xi_M$ is the infinitesimal generator of the action corresponding to $\xi$. In other words, $J$ is a momentum mapping provided
$$X_{\hat{J}(\xi)}=\xi_M,$$
for all $\xi\in \mathfrak{g}$. 
\end{defi}

The next result clarifies the role played by momentum map to explain the relation between symmetries and conserved quantities in Hamiltonian systems.

\begin{prop}\label{conserved} In the hypothesis of \ref{momentummap}, let $J$ be a momentum mapping and let $H:M\rightarrow \R$ be a $G$-invariant function on $M$, that is,
$$H(x)=H(\Phi_g(x)),\quad\;\text{for all}\; x\in M, g\in G .$$
Then $J$ is a conserved quantity of $X_H$. That is, if $F_t$ is the flow of $X_H$,
$$J(x)=J(F_t(x)),\quad\;\text{for all}\; x\in M.$$

\end{prop}

The proof can be found in \cite{abraham2008foundations} (Theorem 4.2.2). 

We introduce now a special class of momentum maps. We say that $J$ is an Ad$^{*}$-equivariant momentum map if
\begin{equation}\label{equivariant}
J(\Phi_g(x))=Ad^*_{g^{-1}}J(x),
\end{equation}
for each $x\in M$ and each $g\in G$. Equivalentely, the following diagram commutes
$$
\xymatrix {
M \ar[rr]^{\Phi_g} \ar[d]_J && M \ar[d]^J \\
\mathfrak{g}^* \ar[rr]_{Ad^*_{g^{-1}}} && \mathfrak{g}^* 
}.
$$

\begin{obs}\label{invariantlevel}
In this case, each level set of a momentum map is invariant under the action of the isotropy group $G_{\mu}=\lbrace g\in G\,|\, Ad^*_{g^{-1}}(\mu)=\mu\rbrace$. Indeed, if $\mu\in \mathfrak{g}^*,\, x\in J^{-1}(\mu)$ and $g\in G_{\mu}$, we have
$$J(\Phi_g(x))=Ad^*_{g^{-1}}J(x)=Ad^*_{g^{-1}}\mu=\mu,$$
thus $\Phi_g(x)\in J^{-1}(\mu)$. This will be essential to perform the reduction. $\hfill\diamondsuit$
\end{obs}

Not every symplectic action has an associated momentum map (due to the fact that not every locally Hamiltonian vector field is globally Hamiltonian), but we can guarantee the existence in some relevant cases. 

\begin{prop}\label{exact symplectic momentum map} Let $M$ be a symplectic manifold with exact symplectic form $\omega=-d\theta$. Let $\Phi$ be a symplectic action on $M$, and suppose that it leaves $\theta$ invariant, that is, $\Phi_g^*\theta=\theta$. Then, $J:M\rightarrow \mathfrak{g}^*$ defined by
$$J(x)\xi=\theta(\xi_M(x)),$$
is an Ad$^*$-equivariant momentum map for the acion.
\end{prop}

\begin{proof} Let $\xi\in \mathfrak{g}$. We have $\hat{J}(\xi)=i_{\xi_M}\theta$. For all $t\in \R$, invariance of $\theta$ gives $\Phi^*_{exp(-t\xi)}\theta=\theta$, and then $\Lie_{\xi_M}\theta=0$. By Cartan's formula,
$$d i_{\xi_M}\theta+i_{\xi_M}d\theta=0,$$
thus,
$$d(\hat{J}(\xi))=d(i_{\xi_M}\theta)=i_{\xi_M}\omega,$$
so $J$ satisfies \eqref{condmomento} and then it is a mumentum map.

For Ad$^*$-equivariant, we should establish \eqref{equivariant}. By definition of $\hat{J}$ and $Ad^*_{g^{-1}}$, this is equivalent to
$$\hat{J}(\xi)(\Phi_g(x))=\hat{J}(Ad_{g^{-1}}\xi)(x),$$
and because $\hat{J}(\xi)=i_{\xi_M}\theta$, this is true only if
$$i_{\xi_M}\theta(\Phi_g(x))=i_{(Ad_{g^{-1}}\xi)_M}(x).$$
But by proposition 4.1.26 \cite{abraham2008foundations}
$$(Ad_{g^{-1}}\xi)_M=\Phi^*_g\xi_M,$$ 
and then
\begin{align*}
i_{(Ad_{g^{-1}}\xi)_M}\theta(x)&=\theta(\Phi^*_g\xi_M(x))=\theta[d\Phi_g(\Phi^*_g\xi_M(x))]\\
&=\theta[d\Phi_g(d\Phi_{g^{-1}}\lbrace\xi_M(\Phi_g(x))\rbrace)]=\theta[\xi_M(\Phi_g(x))]=i_{\xi_M}\theta(\Phi_g(x)),
\end{align*}
where the second equality follows from the invariance of $\theta$ under the action.
\end{proof}

The previous proposition gives a natural way to define a momentum map in the paradigmatic example of symplectic manifold, that is, the contangent bundle. In particular, momentum map is always defined in the Hamiltonian formalism. Let $M$ be a manifold, $G$ a Lie group and $\Phi :G\times M\rightarrow M$ an action of $G$ on $M$. We can construc an action on $T^*M$ by lifting, for each $g\in G$, the diffeomorphism $\Phi_g$. More precisely, we define the \textbf{lifted action} $\Phi^*:G\times T^*M\rightarrow T^*M$ by
$$\Phi^*(g,\alpha)=(T^*\Phi_{g^{-1}})(\alpha),$$
or, equivalently
$$\Phi^*_g=T^*\Phi_{g^{-1}}.$$
We use $g^{-1}$ instead of $g$ for $\Phi^*$ to be an action. By proposition 5.3.2 \cite{de1989methods}, the partial aplications $\Phi^*_g$ preserve both the canonical symplectic form $\omega_M$ and the Liouville form $\lambda_M$. Then we can apply proposition \ref{exact symplectic momentum map} to define an Ad$^*$-equivariant momentum map on $(T^*M,\omega_M)$. In fact, $J:T^*M\rightarrow \mathfrak{g}^*$ given by,
\begin{equation}\label{momentoTM}
J(\alpha)(\xi)=\alpha (\xi_M(x)),
\end{equation}
for $\alpha\in T_xM^*$ and $\xi\in\mathfrak{g}$, is an Ad$^*$-equivariant momentum map for the lifted action $\Phi^*$. The proof can be found in \cite{abraham2008foundations} (corollary 4.2.11).

\begin{ej}[Traslations and linear momentum]
Let $M=\mathbb{R}^n$, $G=\mathbb{R}^n$, and consider the action of $G$ in $\mathbb{R}^n$ by traslations, that is, 
$$\Phi:\mathbb{R}^n\times \mathbb{R}^n\rightarrow \mathbb{R}^n;\; (\textbf{r},\textbf{q})\mapsto \textbf{r}+\textbf{q}.$$
Of course we have $\mathfrak{g}=\mathbb{R}^n$. In order to construct \eqref{momentoTM} we must know $\boldsymbol{\xi}_M$ for all $\boldsymbol{\xi}\in \mathfrak{g}$.

Let $\boldsymbol{\xi}\in \mathbb{R}^n$. The associated left invariant vector field is given by $X_{\boldsymbol{\xi}}(\textbf{r})=d_\textbf{0}L_\textbf{r}(\boldsymbol{\xi})$. Now, since $L_\textbf{r}(\textbf{h})=\textbf{r}+\textbf{h}$, we have $d_\textbf{h}L_\textbf{r}=Id$, and then $X_{\boldsymbol{\xi}}(\textbf{r})=\boldsymbol{\xi}$. We conclude that invariant vector fields are constant.

Hence, the integral curve of $X_{\boldsymbol{\xi}}$  passing through $\textbf{0}$ at $t=0$ is $\varphi_{\boldsymbol{\xi}}(t)=t\boldsymbol{\xi}$. The field $\boldsymbol{\xi}_M$ is then given by,

$$\boldsymbol{\xi}_M(\textbf{q})=\left. \dfrac{\partial}{\partial t}\right|_{t=0} \left(\Phi(exp(t\boldsymbol{\xi}),\textbf{q}) \right)=\left. \dfrac{\partial}{\partial t}\right|_{t=0} \left(t\boldsymbol{\xi}+\textbf{q} \right)=\boldsymbol{\xi}.$$

In coordinates, $\textbf{q}=(q^i)$, $\textbf{p}=(p_i)$, $\boldsymbol{\xi}=(\xi^j)$ and $\boldsymbol{\xi}_M(q^i)=\xi^j\partial/\partial q^j$, thus \eqref{momentoTM} is given by 
$$J(q^i,p_i)(\xi^j)=(p_idq^i)(\xi^j\partial/\partial q^j)=p_i\xi^i,$$
that is, 
$$J(\textbf{q},\textbf{p})(\boldsymbol{\xi})=\textbf{p}\cdot \boldsymbol{\xi}.$$

Identifying $\mathfrak{g}^*$ with $\mathbb{R}^n$ via the dot product, we conclude that $J(\textbf{q},\textbf{p})=\textbf{p}$, that is, the linear momentum. $\hfill\diamondsuit$

\end{ej}

We turn now to the question of reduction. Via the momentum map, reduction can be performed at each level set of the momentum map. Let $(M,\omega)$ be a symplectic manifold, $\Phi:G\times M\rightarrow M$ a symplectic action on $M$ and $J$ an Ad$^*$-equivariant momentum mapping for the action. Let $\mu\in \mathfrak{g}^*$ be a regular value of $J$, and let $G_\mu$ be the isotropy group of $G$ under the co-adjoint action. By Remark \ref{invariantlevel}, the orbit space $J^{-1}(\mu)/G_\mu$ is well defined. Indeed, if $G_\mu$ acts propper and freely on $J^{-1}(\mu)$, then the orbit space is a manifold  \cite{abraham2008foundations} (proposition 4.1.23). We will show that in fact  $J^{-1}(\mu)/G_\mu$ is a symplectic manifold. In the special case where $G_\mu=G$ this can be deduced directly from coisotropic reduction. 
We need the following lemma.

\begin{lema}\label{lemahojaorbita} In the above conditions, if $x\in J^{-1}(\mu)$, then:
\begin{enumerate}
\item[i)]$T_x(G_\mu x)=T_x(Gx)\cap T_xJ^{-1}(\mu)$, where $G_\mu x=\lbrace\Phi_g(x)\,|\,g\in G_\mu\rbrace$ and $Gx=\lbrace\Phi_g(x)\,|\,g\in G\rbrace$ are the orbits of $x$ by the action of $G_\mu$ and $G$, respectively.
\item[ii)]$T_xJ^{-1}(\mu)=T_x(Gx)^\perp$.
\item[iii)] In particular, if $G=G_\mu$, then $T_x(Gx)\subset T_xJ^{-1}(\mu)$ and $J^{-1}(\mu)$ is coisotropic. Moreover
\begin{equation}\label{hojaorbita}
T_xJ^{-1}(\mu)^\perp=T_x(Gx).
\end{equation}
\end{enumerate}
\end{lema} 
\begin{proof}i) Trivially, $G_\mu x\subset Gx$, and $G_\mu x\subset J^{-1}(\mu)$ by remark \ref{invariantlevel}, so we have one inclusion. For the other, let $u\in T_x(Gx)\cap T_xJ^{-1}(\mu)$. By corollary 4.1.22 \cite{abraham2008foundations}, we know that
\begin{equation}\label{orbitaxiM}
T_x(Gx)=\lbrace\xi_M(x)\,|\,\xi\in \mathfrak{g}\rbrace,
\end{equation}
and if $\mathfrak{g}_\mu\subset \mathfrak{g}$ is the lie subalgebra associated to the Lie subgroup $G_\mu\subset G$, we have
$$T_x(G_\mu x)=\lbrace\xi_M(x)\,|\,\xi\in \mathfrak{g}_\mu\rbrace.$$
Since $u\in T_x(Gx)$, there exists $\xi\in \mathfrak{g}$ such that $\xi_M(x)=u$. We must show that in fact $\xi\in \mathfrak{g}_\mu$. Using that $J$ is Ad$^*$-equivariant we get
$$d_xJ(\xi_M(x))=\xi_{\mathfrak{g}^*}(\mu).$$
Indeed,
\begin{align*}
\xi_{\mathfrak{g}^*}(\mu)&=\left.\dfrac{\partial}{\partial t}\right|_{t=0} \left(Ad^*_{exp(t\xi)}J(x) \right)=\left.\dfrac{\partial}{\partial t}\right|_{t=0} \left[J(\Phi(exp(t\xi),x)) \right]\\ &=d_xJ\left(\left.\dfrac{\partial}{\partial t}\right|_{t=0} (\Phi(exp(t\xi),x))\right)=d_xJ(\xi_M(x)).
\end{align*}
Thus, since $u\in T_xJ^{-1}(\mu)$, $\xi_{\mathfrak{g}^*}(\mu)=d_xJ(u)=0$. From this we deduce that $\mu$ is a fixed point of $Ad^*_{exp(-t\xi)}$ and then $exp(\xi)\in G_\mu$. Finally, basic Lie group theory leads to $\xi\in G_\mu$.

ii) Since $J$ is a momentum map, given $\xi\in \mathfrak{g}$ and $v\in T_xM$,
$$\omega(\xi_M(x),v)=d_x\hat{J}(\xi)(v)=d_xJ(v)(\xi).$$
Thus $v\in \ker d_xJ=T_x(J^{-1}(\mu))$ if and only if $\omega(\xi_M(x),v)=0$ for all $\xi \in \mathfrak{g}$, so
$$T_xJ^{-1}(\mu)=\lbrace\xi_M(x)\,|\,\xi\in \mathfrak{g}\rbrace^{\perp}=T_x(Gx)^\perp.$$

iii) By i), $T_x(Gx)\subset T_x(J^{-1}(\mu))$. Using ii), we have
$$T_xJ^{-1}(\mu)^\perp=(T_x(Gx)^\perp)^\perp=T_x(Gx),$$
where the second equality is an easy exercise of linear algebra. Finally, since $T_x(Gx)\subset T_xJ^{-1}(\mu)$, we deduce that $T_xJ^{-1}(\mu)$ is coisotropic.
\end{proof}

Formula \eqref{hojaorbita} says that the orbits under the action of $G$ coincide with the characteristic distribution of $J^{-1}(\mu)$. The following theorem states the symplectic reduction via the momentum map.

\begin{teo}\label{symred} Let $(M,\omega)$ be a symplectic manifold, $\Phi:G\times M\rightarrow M$ a symplectic action on $M$ and $J$ an Ad$^*$-equivariant momentum mapping for the action. Let $\mu\in \mathfrak{g}^*$ be a regular value of $J$ which is a fixed point of $G$ under the coadjoint action, that is, $G_\mu=G$. Suppose that the action of $G$ is free and proper on $J^{-1}(\mu)$. Then, $M_\mu=J^{-1}(\mu)/G$ has a unique symplectic form $\omega_\mu$ such that
\begin{equation}\label{omegamu}
\pi_\mu^*\omega_\mu=i_\mu^*\omega,
\end{equation}
where $\pi_\mu:J^{-1}(\mu)\rightarrow M_\mu$ is the canonical projection and $i_\mu:J^{-1}(\mu)\rightarrow M$ is the inclusion.
\end{teo}
\begin{proof} As we mencioned above, the hypothesis guarantee that $M_\mu$ is a manifold. Using \eqref{hojaorbita}, we get that the quotients $M_\mu$ and $J^{-1}(\mu)/TJ^{-1}(\mu)^\perp$ coincide, and then the result follows from theorem \ref{symcoired}.
\end{proof}

The conclusion of the previous theorem remains true even if the isotropy group $G_\mu$ does not coincide with the whole group $G$. In this case a new proof is needed, since $J^{-1}(\mu)$ is no more coisotropic, and then we can not apply coisotropic reduction. For a proof, see, for example, \cite{abraham2008foundations} (theorem 4.3.1). The following result is proved in the general situation just mentioned.

\begin{teo}\label{SHR}(Symplectic Hamiltonian reduction). Under the assumptions of \ref{symred}, let $H:M\rightarrow \R$ be a $G$-invariant function on $M$. The flow $H_t$ of $X_H$ leaves $J^{-1}(\mu)$ invariant and commutes with the action of $G_\mu$ in $J^{-1}(\mu)$, inducing a flow $H_t^\mu$ on $M_\mu$ satisfying 
\begin{equation}\label{Ht}
\pi_\mu\circ H_t=H_t^\mu\circ \pi_\mu. 
\end{equation}
This flow is Hamiltonian, and the associated Hamiltonian function verifies 
\begin{equation}\label{H}
H_\mu\circ \pi_\mu=H\circ i_\mu.
\end{equation}
The following diagram depicts the situation.
$$
\xymatrix {
M \ar[rd]^{H}\\ J^{-1}(\mu) \ar[u]^{i_\mu} \ar[d]_{\pi_\mu}& \R  \\
M_\mu \ar[ur]_{H_\mu} }
$$
$H_\mu$ is usually called the reduced Hamiltonian.
\end{teo}
\begin{proof} By \ref{conserved}, $J$ is an integral for $X_H$, thus $J^{-1}(\mu)$ is invariant under the flow $H_t$. Let $p\in M_\mu$, $x\in \pi^{-1}_\mu(p)$. Define $H_t^\mu(p)=\pi_\mu(H_t(x))$. This flow trivially satisfies \eqref{Ht} and is well defined since $H_t$ commutes with the action of $G_\mu$. In fact, if $y\in  \pi^{-1}_\mu(p)$, then $y=\Phi_g(x)$ for some $x\in G_\mu$, so
$$\pi_\mu(H_t(y))=\pi_\mu(H_t(\Phi_g(x)))=\pi_\mu(\Phi_g(H_t(x)))=\pi_\mu(H_t(x)).$$
Now set $H_\mu(p)=H(x)$. This is well defined because $H$ is $G$-invariant. Indeed, if $y\in  \pi^{-1}_\mu(p)$,
$$H(y)=H(\Phi_g(x))=H(x).$$
Let $Y$ be the infinitesimal generator of $H_t^\mu$. We claim that $Y=X_{H_\mu}$. Let $[v]=T_x\pi_\mu(v)\in T_pM_\mu$. Then,
%\begin{align*}
%d_pH_\mu[v]&=d_pH_\mu(T_x\pi_\mu(v))=d_x(H_\mu\circ \pi_\mu)(v)=d_x\,(H\circ i_\mu)(v)\\&\,=\,d_xH(d_xi_\mu(v))\;=\;i^*_\mu\, %dH(v)\;=\;i^*_\mu\,\omega(X_H,v)\,=\pi^*_\mu\omega_\mu(X_H,v).
%\end{align*}
\begin{align*}
d_pH_\mu[v]&=d_pH_\mu(T_x\pi_\mu(v))=d_x(H_\mu\circ \pi_\mu)(v)=d_x\,(H\circ i_\mu)(v)=d_xH(T_xi_\mu(v))\\&=i^*_\mu\, dH(v)=i^*_\mu\,\omega(X_H,v)=\pi^*_\mu\omega_\mu(X_H,v)=\omega_\mu(p)(T_x\pi_\mu(X_H),[v]).
\end{align*}
where we have used \eqref{H}, \eqref{Ht} and \eqref{omegamu}. Finally, 
\begin{equation}\label{YXH}
T_x\pi_\mu(X_H)=T_x\pi_\mu\left(\left.\dfrac{\partial}{\partial t}\right|_{t=0}H_t(x)\right)=\left.\dfrac{\partial}{\partial t}\right|_{t=0}(\pi_\mu(H_t(x)))=\left.\dfrac{\partial}{\partial t}\right|_{t=0}(H_t^\mu(p))=Y(p).
\end{equation}
Hence $Y=X_{H_\mu}$, as we claimed. Then $H_t^\mu$ is Hamiltonian with associated function $H_\mu$.
\end{proof}

\begin{obs}(Reconstruction of the dynamics). Hamiltonian reduction allows us to simplify Hamilton equations. Indeed, if the flow $H_t^\mu$ of the reduced space is known, one can recover the flow $H_t$ on $J^{-1}(\mu)$. Let $p_0\in J^{-1}(\mu)$ and consider the integral curve $c(t)$ of $X_H$ such that $c(0)=p_0$. Of course, $c_\mu(t)=\pi_\mu(c(t))$ is an integral curve of $X_{H_\mu}$. The question is whether $c(t)$ can be found in terms of $c_\mu(t)$. 

To do so, take $d(t)\in J^{-1}(\mu)$ a smooth curve such that $\pi_\mu(d(t))=c_\mu(t)$ and $d(0)=p_0$. Then there exists a smooth curve $g(t)$ in $G_\mu\subset G$ such that
\begin{equation}\label{reconstruction}
c(t)=\Phi_{g(t)}d(t).
\end{equation} 
Now, using that $c(t)$ is an integral curve of $X_H$,
\begin{align*}
X_H(c(t))&=\dot{c}(t)\\&=d_{d(t)}\Phi_{g(t)}(\dot{d}(t))+d_{d(t)}\Phi_{g(t)}(d_g(t)L_{g(t)^{-1}}(\dot{c}(t)))_M(d(t)).
\end{align*}
and by the $G$-invariance of $X_H$,
$$X_H(d(t))=\dot{d}(t)+(d_g(t)L_{g(t)^{-1}}(\dot{c}(t)))_M(d(t)),$$
which is an equation for $g(t)$ only in terms of $d(t)$. To solve it, we first consider the algebraic problem,
$$(\xi(t))_M(d(t))=X_H(d(t))-\dot{d}(t),$$
for $\xi(t)\in \mathfrak{g}$, and the we solve
$$\left\{ \begin{array}{l}
             \dot{g}(t)=d_eL_{g(t)}\xi(t), \\
             g(0)=e.
             \end{array}
\right.$$
If we introduce the unique solution of the above problem in \eqref{reconstruction}, we obtain the desired integral curve $c(t)$. $\hfill\diamondsuit$
\end{obs}

\begin{ej}(Angular momentum). Let $M=\mathbb{R}^3$ and consider the action of $G=SO(3)$ by rotations,
$$\Phi: SO(3)\times \mathbb{R}^3\rightarrow \mathbb{R}^3; \; (O,\textbf{q})\mapsto O\textbf{q}.$$
The lifted action is then given by
$$\Phi: SO(3)\times T^*\R^3 \rightarrow T^*\mathbb{R}^3; \; (O,\textbf{q},\textbf{p})\mapsto (O\textbf{q},\textbf{p}O^t).$$
Let $H:T^*M\rightarrow \R$ be a spherically symmetric Hamiltonian function, that is, $H(\textbf{q},\textbf{p})=H(\|\textbf{q}\|,\|\textbf{p}\|)$. It is known that $\mathfrak{g}=\mathfrak{so}(3)=\lbrace o\in\R^{3\times 3}| o^t+o=0 \rbrace$. The identification,
$$\begin{pmatrix}
0 & -x_3 & x_2\\
x_3 & 0 & -x_1\\
-x_2 & x_1 & 0
\end{pmatrix}
\equiv
\begin{pmatrix}
 x_1\\
x_2 \\
x_3
\end{pmatrix},
$$
defines a Lie algebra isomorphism between $\mathfrak{so}(3)$ and $\R^3$ with the cross product. It can be shown that, given $\boldsymbol{\xi}\in \R^3\simeq \mathfrak{so}(3)$, we get $\boldsymbol{\xi}_M(\textbf{q})=\boldsymbol{\xi}\times \textbf{q}$. 

Using coordinates $\textbf{q}=(q^i)$, $\textbf{p}=(p_i)$, $\boldsymbol{\xi}=(\xi^j)$ and $\boldsymbol{\xi}\times \textbf{q}=(\boldsymbol{\xi}\times \textbf{q})^j\partial/\partial q^j$, one has
$$J(q^i,p_i)(\xi^j)=(p_idq^i)((\boldsymbol{\xi}\times \textbf{q})^j\partial/\partial q^j)=p_i(\boldsymbol{\xi}\times \textbf{q})^i,$$
that is,
$$J(\textbf{q},\textbf{p})(\boldsymbol{\xi})=\textbf{p}\cdot (\boldsymbol{\xi}\times \textbf{q})=(\textbf{q}\times \textbf{p})\cdot \boldsymbol{\xi}.$$

Thus identifying $\mathfrak{g}^*$ with $\mathbb{R}^3$ via the dot product, we conclude that $J(\textbf{q},\textbf{p})=\textbf{q}\times \textbf{p}$, that is, the usual angular momentum. Without loss of generality, suppose $\mu=(0,0,\mu_0)$. Then, if $(\textbf{q},\textbf{p})\in J^{-1}(\mu)$, both $\textbf{q}$ and $\textbf{p}$ lie on the $xy$-plane and $q^1p_2-p_1q^2=\mu_0$. Finally, using theorem \ref{SHR}, we obtain a reduced Hamiltonian system $(M_\mu,H_\mu)$ over a $2$-dimensional manifold. 
$\hfill\diamondsuit$
\end{ej}

\label{sec1.3}
\chapter{Reduction in contact geometry}\label{chapter2}
\section{Contact geometry}
In this section we introduce some basic concepts of contact geometry. Some references textbooks are \cite{blair2010riemannian, mcinerney2013first}. However, the main reference for this part is \cite{CHS}.
\begin{defi}(Contact manifold). A contact manifold is a pair $(M,\eta)$, where $M$ is a $2n+1$ dimensional manifold and $\eta$ is a $1$-form on $M$, which is nondegenerate in the following sense: for all $x\in M$,
\begin{equation}\label{nondegeneracy}
\eta\wedge(d\eta)^n(x)\neq 0,
\end{equation}
that is, $\eta\wedge (d\eta)^n$ is a volume form. We will call $\eta$ a contact form.
\end{defi} 

Given a contact manifold $(M,\eta)$, two natural distributions arise:
$$\HH=ker\,\eta,$$
$$\mathscr{V}=ker\;d\eta,$$   
which are called the horizontal and vertical distributions, respectively. $\HH$ is also called the contact distribution.  Notice that $dim(\mathscr{H})=2n$ and that $d\eta|_{\HH}$ is nondegenerate.
%Notice that the horizontal distribution has dimension $2n$.

\begin{obs}
It can be shown that condition \eqref{nondegeneracy} implies that the contact distribution is not integrable. Indeed, some authors \cite{blair2010riemannian, loose2001reduction} define contact manifolds as odd-dimensional manifolds with a contact distribution, that is, a maximally nonintegrable codimension $1$ distribution. Trivially, every contact manifold defines a contact manifold in this wider sense, but the converse is not true. In fact, a distribution in the above hypothesis is always given locally by the kernel of a $1$-form, but not necessarily globally. However, even in the case that the later is true, the corresponding $1$-form is not uniquely defined. 

We will not work with this wider definition, since a fixed contact form is needed to study the dynamics of contact Hamiltonian systems. For instance, correspondence between functions and their contact hamiltonian fields depends crucially on the contact form.  $\hfill\diamondsuit$ 
\end{obs}

There is a Darboux theorem for contact manifolds. Namely, if $(M,\eta)$ is a $2n+1$-dimensional contact manifold, there exists a neighborhood of each point with coordinates $(q^1,...,q^n,p_1,...,p_n,z)$ such that,
$$\eta=dz-p_idq^i,$$
$$d\eta=dq^i\wedge dp_i.$$

Another consequence of condition \eqref{nondegeneracy} is the existence of a unique vector field $\RR$, called Reeb vector field, such that
\begin{equation}\label{Reeb}
i_{\RR} d\eta=0,\quad\;i_{\RR}\eta=1.
\end{equation}
In Darboux coordinates we have,
$$\RR=\frac{\partial }{\partial z}.$$

In analogy with the symplectic case, each contact structure carries a vector bundle isomorphism
\begin{align}
\nonumber \flat:TM &\rightarrow T^*M, \\
v &\mapsto i_vd\eta+\eta(v)\eta.
\end{align}
We denote $\sharp$ to be the inverse of $\flat$. We can also consider $\flat$ as an isomorphism between vector fields and $1$-forms on $M$. Indeed, we have $\flat(\mathscr{R})=\eta$, so $\mathscr{R}$ is the dual object of $\eta$ in this setting.

Finally, consider two contact manifolds $(M,\eta),(N,\mu)$. A diffeomorphism $F:M\rightarrow N$ is said to be a contactomorphism if it preserves the contact form, that is,
$$F^*\mu=\eta.$$

\subsection{Contact Hamiltonian systems}
Next, we use the vector bundle isomorphism just defined to introduce dynamics in contact manifolds.
\begin{defi}(Hamiltonian vector field). Let $H:M\rightarrow \R$ be a smooth map on a contact manifold $(M,\eta)$. We define the associated Hamiltonian vector field by,
\begin{equation}\label{X_Hcontacto}
X_H=\sharp(dH)-(\mathscr{R}(H)+H)\mathscr{R},
\end{equation}
or equivalenty,
\begin{equation}\label{dHcontacto}
\flat(X_H)=dH-(\mathscr{R}(H)+H)\eta.
\end{equation}
\end{defi}
In Darboux coordinates, we get the following expression,
$$X_H=\dfrac {\partial H}{\partial p_i}\dfrac{\partial}{\partial q^i}-\left(\dfrac{\partial H}{\partial q^i}+p_i\dfrac{\partial H}{\partial z}\right)\dfrac{\partial}{\partial p_i}+\left(p_i\dfrac {\partial H}{\partial p_i}-H\right)\dfrac{\partial}{\partial z}.$$
Then, if $(q^i(t),p_i(t),z(t))$ is an integral curve of $X_H$, we have
\begin{align*}
\dfrac{dq^i}{dt}&=\dfrac {\partial H}{\partial p_i},\\ \dfrac{dp_i}{dt}&=-\dfrac{\partial H}{\partial q^i}-p_i\dfrac{\partial H}{\partial z},\\ \dfrac{dz}{dt}&=p_i\dfrac {\partial H}{\partial p_i}-H,
\end{align*}
which are known as the dissipative Hamilton equations. Notice that we recover the conservative Hamilton equations in case $\mathscr{R}(H)=\partial H/\partial z=0$. The above equations are dissipative in the sense that the hamiltonian function $H$ is no longer preserved during the evolution of the system.

\begin{prop} Let $(M,\eta,H)$ be a contact Hamiltonian system. Then $H$ is not preserved by the flow of the Hamiltonian vector field $X_H$. Indeed,
$$\mathscr{L}_{X_H}H=-\mathscr{R}(H)H=-H\dfrac{\partial H}{\partial z},$$
and the contact volume element $\Omega=\eta\wedge(d\eta)^n$ is also not preserved,
\begin{equation}\label{lievolumen}
\mathscr{L}_{X_H}\Omega=-(n+1)\mathscr{R}(H)\Omega.
\end{equation}
\end{prop}

\begin{proof} First we will prove that $\eta(X_H)=-H$. By definition of $\flat$ we have
\begin{equation}\label{bemolsostenido}
dH=\flat(\sharp(dH))=i_{\sharp(dH)}d\eta+\eta(\sharp(dH))\eta,
\end{equation}
so, using Reeb vector field properties \eqref{Reeb}, 
\begin{equation}\label{Reta}
\mathscr{R}(H)=dH(\mathscr{R})=\eta(\sharp(dH)).
\end{equation}
Then, by definition of $X_H$,
\begin{equation}\label{-H}
\eta(X_H)=\eta(\sharp(dH))-\mathscr{R}(H)-H=-H.
\end{equation}
Using this and \eqref{dHcontacto}, energy dissipation follows. In fact,
\begin{align*}
\mathscr{L}_{X_H}H=dH(X_H)&=\flat(X_H)(X_H)+(\mathscr{R}(H)+H)\eta(X_H)\\ &= d\eta(X_H,X_H)+\eta(H)^2+(\mathscr{R}(H)+H)\eta(X_H)\\&=0+H^2-\mathscr{R}(H)H-H^2\\ &=-\mathscr{R}(H)H.
\end{align*}
Now we prove \eqref{lievolumen}. First we will compute $\mathscr{L}_{X_H}\eta$. Using Cartan's formula, \eqref{-H}, and \eqref{X_Hcontacto}, one has
$$\mathscr{L}_{X_H}\eta=i_{X_H}d\eta+di_{X_H}\eta=i_{\sharp dH}d\eta-dH,$$
and combining \eqref{Reeb} and \eqref{Reta}, we get
$$\mathscr{L}_{X_H}\eta=-\mathscr{R}(H)\eta.$$
Finally, the product rule gives
\begin{align*}
\mathscr{L}_{X_H}\eta\wedge(d\eta)^n&=-\mathscr{R}(H)\eta\wedge(d\eta)^n+n\eta\wedge (d\eta)^{n-1}\wedge d\mathscr{L}_{X_H}\eta\\&=-\mathscr{R}(H)\eta\wedge(d\eta)^n+n\eta\wedge (d\eta)^{n-1}\wedge(-(d\mathscr{R}(H))\eta-\mathscr{R}(H)d\eta)\\&=-(n+1)\mathscr{R}(H)\eta\wedge(d\eta)^n.
\end{align*}
as we claimed.
\end{proof}

\begin{ej}(Particle in a potential with friction). Consider a Hamiltonian system given by
$$H(q,p,z)=H_{cm}(q,p)+\gamma z,$$
where $H_{cm}$ is the $1$-dimensional Hamiltonian of a particle in a potential, that is
$$H_{cm}(q,p)=\dfrac{p^2}{2m}+V(q),$$
and $\gamma$ is a constant. The Hamiltonian $H$ discribes a system with dissipation linear in the velocity, with $\gamma$ being the friction parameter. The dynamics of the system is then given by the contact Hamilton equations:
\begin{align*}
\dot{q}&=\dfrac{p}{m},\\ \dot{p}&=-\dfrac{\partial V}{\partial q}-\gamma z,\\ \dot{z}&=\dfrac {p^2}{2m}-V(q)-\gamma z.
\end{align*}
Combining the first two equations, we obtain the equation for the position of the system,
$$\ddot{q}+\gamma \dot{q}+\dfrac{1}{m}\dfrac{\partial V}{\partial q}=0,$$
which is the usual damped Newton equation. We conclude that dissipation is described by the contact formalism.
$\hfill\diamondsuit$
\end{ej}

\subsection{Jacobi manifolds}
Symplectic or contact manifolds are examples of a more general geometric structure, the so-called Jacobi manifolds. 

\begin{defi}(Jacobi manifold). A Jacobi manifold is a triple $(M,\Lambda,E)$, where $M$ is a manifold, $\Lambda$ is $2$-vector (a skew-symmetric contravariant $2$-tensor field) and $E$ is a vector field on $M$ satisfying
\begin{align*}
[\Lambda,\Lambda]&=2E\wedge \Lambda,\\
\Lie_E\Lambda &=[E,\Lambda]=0,
\end{align*}
where $[\cdot,\cdot]$ is the Schouten-Nijenhius bracket, \cite{ nijenhuis, Schouten1953OnTD}.
\end{defi}

Symplectic and contact manifolds can be understood as Jacobi manifolds. 
In fact, if $(M,\omega)$ is a symplectic manifold, and $\sharp_\omega:T^*M\rightarrow TM$ is the induced vector bundle isomorphism, we define a Jacobi structure by taking,
$$\Lambda(\alpha,\beta)=\omega(\sharp_\omega(\alpha),\sharp_\omega(\beta)), \quad E=0.$$
Now, if $(M,\eta)$ is a contact manifold, and $\sharp$ is the induced vector bundle isomorphism, we can take
$$\Lambda(\alpha,\beta)=-d\eta(\sharp(\alpha),\sharp(\beta)), \quad E=-\RR,$$
and we obtain a Jacobi structure on $M$.

The Jacobi strucutre induces a vector bundle morphism between covectors and vectors,
\begin{align*}
\sharp_\Lambda:TM&\rightarrow T^*M,\\ \alpha&\mapsto \Lambda(\alpha,\cdot),
\end{align*}
which, as always, can be considered as a $\mathscr{C}^\infty(M)$-modules morphism between $1$-forms and vector fields on $M$.

If $(M,\eta)$ is a contact manifold, $\sharp_\Lambda$ can be written
$$\sharp_\Lambda(\alpha)=\sharp(\alpha)-\alpha(\mathscr{R})\mathscr{R}.$$

\begin{prop}\label{imsharp} For a contact manifold, $\sharp_\Lambda$ is not an isomorphism. In fact, $\ker \sharp_\Lambda=\langle\eta\rangle$ and im $ \sharp_\Lambda=\mathscr{H}$.
\end{prop}
\begin{proof} We proceed with the first claim. Using Reeb vector field properties, one has
$$\sharp_\Lambda(\eta)=\sharp(\eta)-\eta(\mathscr{R})\RR=\RR-\RR=0,$$
thus $\langle \eta\rangle\subset \ker \sharp_\Lambda$. Now, if $\alpha\in \ker \sharp_\Lambda$, then
$$\sharp_\Lambda(\alpha)=\sharp(\alpha)-\alpha(\RR)\RR=0,$$
and applying $\flat$,
$$\flat(\sharp(\alpha))-\alpha(\RR)\flat(\RR)=\alpha-\alpha(\RR)\eta=0,$$
hence $\alpha=\alpha(\RR)\eta\in \langle \eta\rangle$ and then $\ker \sharp_\Lambda=\langle\eta\rangle$ and $\sharp_\Lambda$ is not an isomorphism.
For proving the second claim, take $\alpha$ $1$-form in $M$. Using \eqref{Reta} with $\alpha$ instead of $dH$ gives
$$\alpha(\RR)=\eta(\sharp(\alpha)),$$
and then
$$\eta(\sharp_\Lambda(\alpha))=\eta(\sharp(\alpha))-\alpha(\RR)=0,$$
hence im $\sharp_\Lambda\subset \mathscr{H}$. For the other inclusion, take $v\in \mathscr{H}$. We will show that $\alpha=i_vd\eta$ verifies $\sharp_\Lambda(\alpha)=v$. In fact,
$$\flat(v)=i_vd\eta+\eta(v)\eta=i_vd\eta+0=\alpha,$$
so $\sharp(\alpha)=\flat^{-1}(\alpha)=v$. Therefore,
$$\sharp_\Lambda(v)=\sharp(v)-\alpha(\RR)\RR=v-i_vd\eta(\RR)\RR=v,$$
as we wanted to show. We conclude that im $ \sharp_\Lambda=\mathscr{H}$.
\end{proof}
 
\subsection{Submanifolds of a contact manifold}
As in the case of symplectic manifolds, it is worth studiyng some relevant types of submanifolds of contact manifolds \cite{CHS, esen2021contact}. The definition is slightly different from the symplectic one, as it uses the Jacobi inherited morphism $\sharp_\Lambda$ instead of the isomorphism $\sharp$.
\begin{defi} Let $(M,\eta)$ be a contact manifold, $x\in M$, and $\Delta_x$ a linear subspace of $T_xM$. Then the subspace,
$$\Delta_x^{\perp_\Lambda}=\sharp_{\Lambda}(\Delta_x^\circ)$$
is called the contact complement of $\Delta_x$ in $T_xM$. Here $\Delta_x^\circ=\lbrace\alpha\in T_xM^*\,|\,\alpha(\Delta_x)=0\rbrace$ is the annihilator of $\Delta_x$.
\end{defi}
The above definition can be extented to distributions $\Delta\subset TM$ and submanifolds $N\subset M$ by taking the complement pointwise in each tangent space.

\begin{defi} A submanifold $N$ of a contact manifold $(M,\eta)$ is called,
\begin{itemize}
\item[i)] isotropic if $TN\subset TN^{\perp_\Lambda},$
\item[ii)] coisotropic if $TN^{\perp_\Lambda}\subset TN,$
\item[iii)] Legendrian if $TN=TN^{\perp_\Lambda}.$
\end{itemize}
\end{defi}

Using the $2$-form $d\eta$ we can define another orthogonal complement, just as we did in the symplectic case. Let $\Delta\subset TM$ be a distribution. We define,
$$\Delta^{\perp_{d\eta}}=\lbrace v\in TM\,|\,d\eta(v,\Delta)=0\rbrace.$$
The fact that $(\mathscr{H},d\eta|_{\mathscr{H}})$ is a symplectic space will have consequences on this symplectic like complement. Also, the contact complement is tightly related with it.

\begin{prop}\label{deltararo} Let $(M,\eta)$ be a $2n+1$-dimensional contact manifold and let $\Delta\subset TM$ be a rank-$k$ distribution. Then we have
\begin{equation}\label{subseteq}
\Delta^{\perp_{d\eta}}\cap \mathscr{H}\subseteq \Delta^{\perp_\Lambda}.
\end{equation}
Furthermore, if $\Delta$ is horizontal (that is $\Delta\subset \HH$) or if $\RR\in \Delta$,  the equality holds.
\end{prop}
\begin{proof} Let $v\in\Delta^{\perp_{d\eta}}\cap \mathscr{H}$. By proposition \ref{imsharp}, we know that $\alpha=i_vd\eta$ verifies $\sharp_\Lambda(\alpha)=v$. Now, given $u\in \Delta$, we have
$$\alpha(u)=i_vd\eta(u)=d\eta(v,u)=0,$$
since $v\in \Delta^{\perp_{d\eta}}$. We conclude that $\alpha\in \Delta^\circ$ and so $\Delta^{\perp_{d\eta}}\cap \mathscr{H}\subseteq \Delta^{\perp_\Lambda}$.

For the second claim we argue with dimensions. Note that
$$\Delta^{\perp_{d\eta}}\cap \mathscr{H}=\lbrace v\in \HH\,|\,d\eta(v,\Delta)=0\rbrace.$$
and since $(\mathscr{H},d\eta|_{\mathscr{H}})$ is symplectic and $\Delta\subset \HH$, we get 
$$\dim \Delta^{\perp_{d\eta}}\cap \mathscr{H}=2n-k.$$ 
Note also that $\Delta\subset \HH$ implies $\eta\in \Delta^\circ$. Now, $\dim\Delta^\circ=2n+1-k$ and $ \Delta^{\perp_\Lambda}=\sharp_\Lambda(\Delta^\circ)$, so by proposition \eqref{imsharp} we obtain 
$$\dim \Delta^{\perp_\Lambda}=\dim \Delta^\circ-\dim \ker( \sharp_\Lambda|_{\Delta^\circ})=2n+1-k-1=2n-k,$$
thus dimensions coincide and then the equality holds.

Suposse now that $\RR\in \Delta$. We can take a basis $\lbrace u_1,...,u_{k-1},\RR\rbrace$ for $\Delta$, where $u_i\in \HH$ for $i=1,...,k-1$. By the linearity of $d\eta$,
 $$\Delta^{\perp_{d\eta}}\cap \mathscr{H}=\lbrace v\in \HH\,|\,d\eta(v,\Delta)=0\rbrace=\lbrace v\in \HH\,|\,d\eta(v,u_i)=0,\,\text{for}\, i=1,...,k-1\rbrace.$$
And using that $(\mathscr{H},d\eta|_{\mathscr{H}})$ is symplectic, we have
$$\dim \Delta^{\perp_{d\eta}}\cap \mathscr{H}=2n-(k-1)=2n-k+1.$$ 
But in the case when $\RR\in \Delta$, $\langle \eta \rangle \notin \Delta^\circ$, thus
$$\dim \Delta^{\perp_\Lambda}=\dim \Delta^\circ-\dim \ker( \sharp_\Lambda|_{\Delta^\circ})=2n+1-k,$$
which completes the proof.
\end{proof}

We next classify distributions according to their relative position with the horizontal and vertical distributions.

\begin{defi}\label{posicionrelativa} Let $\Delta\subset TM$ be a rank $k$ distribution. A point $x\in M$ is said to be,
\begin{itemize}
\item[i)] horizontal if $\Delta_x=\Delta_x\cap\HH_x$,
\item[ii)] vertical if $\Delta_x=(\Delta_x\cap\HH_x)\oplus \langle\RR_x\rangle$,
\item[iii)] oblique if $\Delta_x=\Delta_x\cap\HH_x\oplus \langle\RR_x+v_x\rangle$, with $u_x\in \HH_x\backslash \Delta_x$.
\end{itemize}
\end{defi} 

A point $x$ in a submanifold $N\subset M$ is called horizontal, vertical, or oblique, respectively, if $T_xN$ is horizontal, vertical, or oblique.

\begin{obs}\label{dimensioncomplement} (Dimension of the contact complement). If $x\in M$ is either an horizontal or a vertical point, last proposition gives
\begin{equation}\label{igualdad}
\Delta_x^{\perp_{d\eta}}\cap \mathscr{H}_x=\Delta_x^{\perp_\Lambda}.
\end{equation}
In the corresponding proof we showed that $\dim \Delta_x^{\perp_\Lambda}=2n-k$ in the horizontal case, and that $\dim \Delta_x^{\perp_\Lambda}=2n+1-k$ in the vertical one. 
Suppose now that $x$ is oblique. Clearly, $\eta(v_x+\RR_x)=1$ so that $\langle \eta_x \rangle \notin \Delta_x^\circ$, and then
$$\dim \Delta_x^{\perp_\Lambda}=\dim \Delta_x^\circ-\dim \ker( \sharp_\Lambda|_{\Delta_x^\circ})=2n+1-k,$$
that is, $2n+1-k$ is the dimension for oblique points. Furthermore, we can easily prove that the inclusion \eqref{subseteq} is strict in the oblique case. Indeed, we can take a basis $\lbrace u_1,...,u_{k-1},u_k+\RR\rbrace$ for $\Delta_x$, where $u_i\in \HH_x$ for $i=1,...,k-1$, and then
$$\Delta_x^{\perp_{d\eta}}\cap \mathscr{H}_x=\lbrace v\in \HH_x\,|\,d\eta(v,\Delta_x)=0\rbrace=\lbrace v\in \HH_x\,|\,d\eta(v,u_i)=0,\,\text{for}\, i=1,...,k\rbrace.$$
And using that $(\mathscr{H},d\eta|_{\mathscr{H}})$ is symplectic, we have
$$\dim \Delta_x^{\perp_{d\eta}}\cap \mathscr{H}_x=2n-k.$$
Thus dimension do not coincide and the inclusion is strict. $\hfill\diamondsuit$
\end{obs}

The last result of the section is the contact counterpart of proposition \ref{nsymplectic}.
\begin{prop}\label{ncontact} Let $(M,\eta)$ be a $2n+1$-dimensional contact manifold. A submanifold $i:N\hookrightarrow M$ is isotropic if and only if it is horizontal, that is, $\eta|_{TN}=0$. Furthermore, $N$ is Legendrian if and only if it is isotropic and $\dim N=n$. 
\end{prop}
\begin{proof} Fix $x\in N$. If $N$ is isotropic, then $T_xN\subset T_xN^{\perp_\Lambda}=\sharp_\Lambda(T_xN^\circ)$. But by \eqref{imsharp}, im$\sharp_\Lambda=\mathscr{H}$, thus $T_xN\subset \mathscr{H}$ and so $\eta|_{TN}=0$.

For the converse, notice that $i^*\eta=\eta|_{TN}=0$, so
$$i^*d\eta=d(i^*\eta)=0,$$
and then 
$$i^*d\eta(u,v)=d\eta(u,v)=0,$$
for all $u,v\in T_xN$. We conclude that $u\in(T_xN)^{\perp_{d\eta}}$. By hipothesis $u\in \mathscr{H}$, and using proposition \ref{deltararo}, also $u\in (T_xN)^{\perp_\Lambda}$, thus $T_xN\subset(T_xN)^{\perp_\Lambda}$ and $N$ is isotropic.

For the second assertion, if $N$ is Legendrian, then it is isotropic and $T_xN=(T_xN)^{\perp_\Lambda}$. Using the first claim, $N$ is horizontal, so we have
$$\dim T_xN=\dim (T_xN)^{\perp_\Lambda}=2n-\dim T_xN,$$
so that $\dim T_xN=n$, as we wanted to show. 

Finally, if $N$ is isotropic, we have $T_xN\subset(T_xN)^{\perp_\Lambda}$. But $N$ is horizontal and $\dim N=n$, so
$$\dim (T_xN)^{\perp_\Lambda}=2n-n=n.$$
Clearly $T_xN=(T_xN)^{\perp_\Lambda}$ and consequently $N$ is Legendrian.
\end{proof}
\label{sec2.1}
\section{Coisotropic reduction in contact geometry}
In this section, we study the contact analogous of the symplectic coisotropic reduction theorem. Before presenting the theorem, we analyze the structure of the contact complement of coisotropic submanifolds. 
\begin{prop} Let $i:N\hookrightarrow M$ be a coisotropic submanifold, and suppose that $N$ does not have oblique points. We define
$$\eta_0=i^*\eta=\eta|_{TN},$$
$$d\eta_0=i^*(d\eta)=d(i^*\eta).$$
Then,
$$TN^{\perp_\Lambda}=\ker(\eta_0)\cap \ker(d\eta_0).$$
We call $TN^{\perp_\Lambda}$ the characteristic distribution of $N$.
\end{prop}
\begin{proof}\label{conflicto} By equation \eqref{igualdad},
$$TN^{\perp_\Lambda}=TN^{\perp_{d\eta}}\cap \HH=TN^{\perp_{d\eta}}\cap \ker\eta.$$
But $N$ is coisotropic, so $TN^{\perp_\Lambda}\subset TN$, and then
$$TN^{\perp_\Lambda}=(TN^{\perp_{d\eta}}\cap TN)\cap (\ker\eta\cap TN).$$
Clearly, $\ker\eta\cap TN=\ker\eta_0$, and
$$TN^{\perp_{d\eta}}\cap TN=\lbrace v\in TN\,|\,d\eta(v,TN)=0\rbrace=\lbrace v\in TN\,|\,d\eta_0(v,TN)=0\rbrace=ker (d\eta_0),$$
so we get what we wanted.
\end{proof}

Next we present the central result of the section. 
\begin{teo}\label{concoired}(Coisotropic reduction in contact geometry). Let $i:N\hookrightarrow M$ be a coisotropic submanifold which consists of vertical points. Then, $TN^{\perp_\Lambda}$ is an involutive distribution on $N$.

Assume that the leaf space obtained by Frobenius theorem, denoted by $\tilde{N}=N/TN^{\perp_\Lambda}$, is a manifold. 
%and that $N$ does not have horizontal points. 
Let $\pi:N\rightarrow \tilde{N}$ be the projection. Then, there exists a unique $1$-form $\tilde{\eta}\in \Omega^1(\tilde{N})$ such that,
\begin{equation}\label{conred1}
\pi^*\tilde{\eta}=i^*\eta \coloneqq\eta_0.
\end{equation}
Furthermore, $\tilde{\eta}$ is a contact form and $(\tilde{N},\tilde{\omega})$ is a contact manifold.
\end{teo}

\begin{proof}
First, we will show that $TN^{\perp_\Lambda}$ is an involutive distribution. Let $X_1,X_2\in TN^{\perp_\Lambda}$. We have to prove that $\left[X_1,X_2\right]$ takes values in $TN^{\perp_\Lambda}$ as well. Since $TN^{\perp_\Lambda}=\ker(\eta_0)\cap \ker(d\eta_0)$, then $\eta_0(X_1)=\eta_0(X_2)=1$ and $d\eta_0(X_i,Y)=0$ for all vector fields $Y\in \chi(N)$. Hence, by proposition 1.12.3 \cite{de1989methods}
$$0=d\eta_0(X_1,X_2)=X_1(\eta_0(X_2))-X_2(\eta_0(X_1))-\eta_0([X_1,X_2])=-\eta_0([X_1,X_2]),$$
thus $[X_1,X_2]\in \ker(\eta_0)$. Using now the proposition with $d\eta_0$, we have,
\begin{align*}
0&=dd\eta_0(X_1,X_2,Y)=X_1(d\eta_0(X_2,Y))-X_2(d\eta_0(X_1,Y))+Y(d\eta_0(X_1,X_2))
\\ &-d\eta_0(\left[X_1,X_2\right],Y)+d\eta_0(\left[X_1,Y\right],X_2)-d\eta_0(\left[X_2,Y\right],X_1)=-d\eta_0(\left[X_1,X_2\right],Y).
\end{align*}
So we have $[X_1,X_2]\in \ker(d\eta_0)$ and then $[X_1,X_2]\in TN^{\perp_\Lambda}$.
\\

Now, we will prove that there is a unique $1$-form verifying \eqref{conred1}. For the uniqueness, let $p\in \tilde{N}$, $\tilde{u}\in T_p\tilde{N}$,  and let $x\in \pi^{-1}(p)$ and $u\in T_xN$ (which exists because $\pi$ is a submersion) such that $T_x\pi(u)=\tilde{u}$. Suposse there is some $\tilde{\eta}$ such that $\pi^*\tilde{\eta}=\eta_0$.  Then,
\begin{equation}\label{tildeeta}
\tilde{\eta}(p)(\tilde{u})=\tilde{\eta}(\pi(x))(T_x\pi(u))=\pi^*\tilde{\eta}(x)(u)=\eta_0(x)(u).
\end{equation}
The right side does not depend on the $1$-form chosen, and then the $1$-form is unique in case it exists.
\\

For the existence, we will show that \eqref{tildeeta} defines the desired $1$-form. First, we must prove the independence on the chosen $x,u$. 

Fix $x\in \pi^{-1}(p)$. The independence in $T_xN$ follows from $\ker(T_x\pi)=(T_xN)^{\perp_\Lambda}\subset \ker(\eta_0)$. Indeed, given $\bar{u}$ such that $T_x\pi(\bar{u})=\tilde{u}$, we have $u-\bar{u}\in \ker(T_x\pi)\subset \ker(\eta_0)$. Then,
$$\eta_0(x)(\bar{u})=\eta_0(x)(\bar{u})+\eta_0(x)(u-\bar{u})=\eta_0(x)(u).$$
Thus $\tilde\eta(p)$ does not depend on the chosen element of the preimage of $T_x\pi$.

Now, we prove the independence on the point on the leaf chosen. Let $X$ be tangent to the leaf $\pi^{-1}(p)$, that is, $X\in TN^{\perp_\Lambda}$ so $i_X\eta_0=0$ and $i_Xd\eta_0=0$. Thus, by Cartan's formula,
$$\Lie_X\eta_0=di_x\eta_0+i_xd\eta_0=0.$$
We conclude that $\tilde{\eta}$ in \eqref{tildeeta}  is well defined.

Next we prove that $\tilde{\eta}$ is a $1$-form. Since $\pi$ is a submersion, there exists locally a smooth section $\mu:\tilde{N}\rightarrow N$ (that is, $\pi \circ \mu=id_{\tilde{N}}$). Take $\bar{\eta}=\mu^*\eta_0$. The independence in $\pi^{-1}(p)$ just checked gives $\pi^*\bar{\eta}=\eta_0$, and then \eqref{tildeeta} implies $\tilde{\eta}=\bar{\eta}$ wherever $\bar{\eta}$ is defined. This can be made in a neighborhood of each point, thus $\tilde{\eta}$ is a $1$-form. 

%As $\tilde{\eta}$ satisfies \eqref{conred1} by definition, only last to prove that it is a contact form. Let $p\in \tilde{N}$ and let $x\in \pi^{-1}(p)$ be an oblique point. Then there exists $v\in \HH_x\backslash T_xN$ such that $\RR(x)+v\in T_xN$, thus
%$$\tilde{\eta}(p)(T_x\pi(\RR(x)+v))=\eta(x)(\RR(x))+\eta(x)(v)=1,$$
%hence $\tilde{\eta}$ is nondegenerate. Suppose now that $x$ is vertical (there is no more option since $x$ is not horizontal by hypothesis). Then $\RR(x)\in T_xN$ and so $\tilde{\eta}$ is again nondegenerate. On the other hand, $d\tilde{\eta}$ is trivially nondegenerate, since we have taken the quotient by $\ker(\eta_0)\cap\ker(d\eta_0)$. 

As $\tilde{\eta}$ satisfies \eqref{conred1} by definition, only last to prove that it is a contact form. Let $p\in \tilde{N}$ and let $x\in \pi^{-1}(p)$ be a vertical point. Then $\RR(x)\in T_xN$ and so $\tilde{\eta}$ is nondegenerate. On the other hand, $d\tilde{\eta}$ is trivially nondegenerate, since we have taken the quotient by $\ker(\eta_0)\cap\ker(d\eta_0)$.
\end{proof}

As a consequence of last theorem, we have that the projection of a Legendrian submanifold is also a Legendrian submanifold. 

\begin{prop}
Let $L$ be a Legendrian submanifold, $N$ a coistropic submanifold which consists of vertical points, and $\pi$ the asociated projection to the leaf space. If $L$ and $N$ have clean intersection, then the projection $\tilde{L}=\pi(L\cap N)\subset \tilde{N}$ is also a Legendrian submanifold.
\end{prop}
\begin{proof}
Let $2n+1$ be the dimension of $M$, and $n+k+1$ the dimension $N$. Since $L$ is Legendrian, $dim\, L=n$, and using that $N$ dose not have horizontal points, $dim\, (TN^{\perp_\Lambda})=n-k$. Hence,
$$dim \,\tilde{N}=dim\, N-dim (TN^{\perp_\Lambda})=2k+1.$$
By proposition \ref{ncontact}, it suffices to prove that $\tilde{L}$ is horizontal and that $dim\,\tilde{L}=k$.
\\

We proceed with the first claim. Let $[x]\in \tilde{L}$ and $\tilde{u}\in T_{[x]}\tilde{L}$. Since $\pi$ is a submersion, there exists $u\in T_x(N\cap L)$ such that $\tilde{u}=T_x\pi(u)$, and then,
$$\tilde{\eta}([x])(\tilde{u})=\tilde{\eta}(\pi(x))(T_x\pi(u))=\pi^* \tilde{\eta}(x)(u)=\eta (x)(u)=0,$$
since $\pi^*\tilde{\eta}=\eta$ and $L$ is Legendrian, so horizontal.
\\

Now, to prove that $dim\,\tilde{L}=k$, note that,
\begin{equation}\label{abba}
dim\,(N\cap L)+dim\,(N\cap L)^{\perp_\Lambda}=2n,
\end{equation}
since $L$ is horizontal and so $N\cap L$. By equation \eqref{dedemorgan},
$$(N\cap L)^{\perp_\Lambda}=N^{\perp_\Lambda}+L^{\perp_\Lambda}=N^{\perp_\Lambda}+L,$$
and using Grassmann formula, 
$$dim\, (N^{\perp_\Lambda}+L)=dim\, N^{\perp_\Lambda}+dim\, L-dim\,(N\cap L^{\perp_\Lambda})=2n-k-dim\,(N\cap L^{\perp_\Lambda}).$$
Substituting in \eqref{abba}, we get,
$$dim\,(N\cap L)=2n-dim\,(N\cap L)^{\perp_\Lambda}=k+dim\,(N\cap L^{\perp_\Lambda}).$$
We conclude that, 
$$dim\,\tilde{L}=dim\,(N\cap L)-dim\,(N\cap L^{\perp_\Lambda})=k.$$

\end{proof}\label{sec2.2}
\section{Contact momentum map}
In this section, we introduce momentum maps in the context of contact Hamiltonian systems. The procedure is slightly different from the symplectic one. There, we defined momentum map in a general way and then we proved that exact symplectic manifolds carry a natural equivariant momemtum map. Now, since contact manifolds are exact in some sense, we will start with a specific momentum map, and then we will prove that it satisfies the usual momentum map properties.
\begin{defi}\label{contact momentum def}(Contact momentum map). Let $(M,\eta)$ be a connected contact manifold, $G$ a Lie group, and $\Phi:G\times M\rightarrow M$ a contact action of $G$ on $M$; that is, such that the map $\Phi_g:M\rightarrow M$ is a contactomorphism for all $g\in G$. We define the contact mometum map $J:M\rightarrow \mathfrak{g}^*$ by
\begin{equation}\label{contact momentum formula}
J(x)\xi=-\eta(\xi_M(x)),
\end{equation}
where $\xi_M$ is the infinitesimal generator of the action corresponding to $\xi$, that is,
$$\xi_M(x)=\left.\dfrac{\partial}{\partial t}\right|_{t=0} (\Phi(exp(t\xi),x)).$$
\end{defi} 
\begin{prop} Under the assumptions of \ref{contact momentum def}, let $\hat{J}(\xi):M\rightarrow \R$ be defined by
$$\hat{J}(\xi)(x)=J(x)\xi.$$
Then the usual momentum condition is satisfied:
$$d\hat{J}(\xi)=i_{\xi_M}d\eta,$$
or equivalenty
$$X_{\hat{J}(\xi)}=\xi_M.$$
Furthermore, $J$ is equivariant under the coadjoint action.
\end{prop}
\begin{proof} The proof is a minor adaptation of the proof of Proposition \ref{exact symplectic momentum map}.
\end{proof}
Next we consider the question of reduction, just as we did in the symplectic case. Let $(M,\eta)$ be a contact manifold, $\Phi:G\times M\rightarrow M$ a contact action on $M$ and $J$ the associated momentum map. Let $\mu\in \mathfrak{g}^*$ be a regular value of $J$, and let $G_\mu$ be the isotropy group of $G$. Since $J$ is Ad$^*$-equivariant, we know that the orbit space $M_\mu=J^{-1}(\mu)/G_\mu$ is well defined. Indeed, if $G_\mu$ acts propper and freely on $J^{-1}(\mu)$, then the orbit space is a manifold. We will show that in fact $M_\mu$ is a contact manifold. First we need the adaptation of lemma \ref{lemahojaorbita}.

\begin{lema} \label{lemahojaorbitacontacto}In the above conditions, if $x\in J^{-1}(\mu)$, then:
\begin{enumerate}
\item[i)]$T_x(G_\mu x)=T_x(Gx)\cap T_xJ^{-1}(\mu)$, where $G_\mu x, Gx$ are the orbits of $x$ by the action of $G_\mu$ and $G$, respectively.
\item[ii)]$T_xJ^{-1}(\mu)=T_x(Gx)^{\perp_{d\eta}}$.
\item[iii)] In particular, if $G=G_\mu$, then $T_x(Gx)\subset T_xJ^{-1}(\mu)$ and $J^{-1}(\mu)$ is coisotropic and consists of vertical points. Moreover
\begin{equation}\label{hojaorbitacontacto}
T_xJ^{-1}(\mu)^{\perp_\Lambda}=T_x(Gx).
\end{equation}
\end{enumerate}
\end{lema} 
\begin{proof} The proof of i) and ii) is essentially identical to the proof of lemma \ref{lemahojaorbita}. For iii), notice that $\RR\in T_x(Gx)^{\perp_{d\eta}}$ by direct application of Reeb vector field properties. Then, ii) implies that $T_xJ^{-1}(\mu)$ consits of vertical points. Finally, \eqref{hojaorbitacontacto} is a consequence of ii) and lemma \ref{deltararo}.
\end{proof}

Formula \eqref{hojaorbitacontacto} says that the orbits under the action of $G$ coincide with the leaves of the characteristic distribution of $J^{-1}(\mu)$. The following theorem states the contact reduction via the momentum map.

\begin{teo}\label{conred} Let $(M,\eta)$ be a contact manifold, $\Phi:G\times M\rightarrow M$ a contact action on $M$ and $J$ the associated Ad$^*$-equivariant momentum mapping. Let $\mu\in \mathfrak{g}^*$ be a regular value of $J$ which is a fixed point of $G$ under the coadjoint action, that is, $G_\mu=G$. Suppose that the action of $G$ is free and proper on $J^{-1}(\mu)$. Then, $M_\mu=J^{-1}(\mu)/G$ has a unique contact form $\eta_\mu$ such that
\begin{equation}\label{etamu}
\pi_\mu^*\eta_\mu=i_\mu^*\eta,
\end{equation}
where $\pi_\mu:J^{-1}(\mu)\rightarrow M_\mu$ is the canonical projection and $i_\mu:J^{-1}(\mu)\rightarrow M$ is the inclusion. Moreover, the Reeb vector field $\RR_\mu$ of $M_\mu$ is the projection of the Reeb vector field $\RR$ of $(M,\eta)$, that is,
\begin{equation}\label{relacionreebs}
\RR_\mu(p)=T_x\pi_\mu(\RR(x)),
\end{equation}
for all $p\in M_\mu$, $x\in \pi_\mu^{-1}(p)$.
\end{teo}
\begin{proof} The proof of the first part follows mutatis mutandis that of Theorem \ref{symred}. For the second part, let $p\in M_\mu$, $x\in \pi^{-1}_\mu(p)$ and define the following vector field on $M_\mu$,
$$X(p)=T_x\pi_\mu(\RR(x)).$$
We will show that $X$ satisfies the Reeb vector field properties. Then, using the uniquess of the Reeb vector field, we get $X=\RR_\mu$, and the proof is finished. Indeed,
$$\eta_\mu(X(p))=\eta_\mu(T_x\pi_\mu(\RR(x)))=\pi^*_\mu(\RR(x))=\eta(\RR(x))=1,$$
and using that $\pi^*_\mu d\eta_\mu=i^*_\mu d\eta$, we obtain,
$$d\eta_\mu(X(p),\cdot)=d\eta_\mu(T_x\pi_\mu(\RR(x)),\cdot)=\pi^*_\mu d\eta_\mu(\RR(x),\cdot)=d\eta(\RR(x),\cdot)=0,$$
thus $\eta(X)=1$ and $i_Xd\eta=0$, so $X=\RR_\mu$.
\end{proof}

So far, the contact manifold has been reduced. Now we analize the reduction of the dynamics, that is, the reduction of the whole contact Hamiltonian system. The following theorem has the same structure as theorem \ref{SHR}, but the proof is rather different, since Hamiltonian vector fields have distinct definitions in the symplectic and contact formalisms. 

\begin{teo}(Contact Hamiltonian reduction). Under the assumptions of \ref{conred}, let $H:M\rightarrow \R$ be a $G$-invariant function on $M$. The flow $H_t$ of $X_H$ leaves $J^{-1}(\mu)$ invariant and commutes with the action of $G_\mu$ in $J^{-1}(\mu)$, inducing a flow $H_t^\mu$ on $M_\mu$ satisfying 
\begin{equation}\label{Ht2}
\pi_\mu\circ H_t=H_t^\mu\circ \pi_\mu. 
\end{equation}
This flow is Hamiltonian, and the associated Hamiltonian function verifies 
\begin{equation}\label{H2}
H_\mu\circ \pi_\mu=H\circ i_\mu.
\end{equation}
The following diagram depicts the situation.
$$
\xymatrix {
M \ar[rd]^{H}\\ J^{-1}(\mu) \ar[u]^{i_\mu} \ar[d]_{\pi_\mu}& \R  \\
M_\mu \ar[ur]_{H_\mu} }
$$
$H_\mu$ is usually called the contact reduced Hamiltonian.
\end{teo}
\begin{proof} First of all, we will prove that the flow $H_t$ of $X_H$ leaves $J^{-1}(\mu)$ invariant. It suffices to show that $X_H|_{J^{-1}(\mu)}\in \chi(J^{-1}(\mu))$. Since $H$ is $G$-invariant, one has
$$dH(\xi_M)=\xi_M(H)=0,$$
and then by \eqref{orbitaxiM}, $dH\in T_x(Gx)^\circ$. Clearly, $\sharp_\Lambda(dH)\in\sharp_\Lambda(T_x(Gx)^\circ)=T_x(Gx)^{\perp_\Lambda}$. Now let $x\in J^{-1}(\mu)$. By definition of $X_H$,
$$X_H(x)=\sharp_\Lambda(d_xH)-H(x)\RR(x)\subset T_x(Gx)^{\perp_\Lambda}\oplus \mathscr{V}=T_x(Gx)^{\perp_{d\eta}}=T_xJ^{-1}(\mu).$$
where we have used \ref{lemahojaorbitacontacto} ii). We conclude that $X_H|_{J^{-1}(\mu)}\in \chi(J^{-1}(\mu))$. Now the flow on \eqref{Ht2} is trivially well defined. 

For the second part, let $Y$ be the infinitesimal generator of $H_t^\mu$. We claim that $Y=X_{H_\mu}$. We must show that
\begin{equation}\label{relacionYHmu}
\flat(Y)=dH_\mu-(\RR_\mu(H_\mu)+H_\mu)\eta_\mu.
\end{equation}
Let $p\in M_\mu$, $x\in \pi^{-1}_\mu$, and $[v]=T_x\pi_\mu(v)\in T_pM_\mu$. From \eqref{YXH} we know that,
$$Y(p)=T_x\pi_\mu(X_H(x)).$$
Then, by definition of $\flat$ and using $\pi_\mu^*\eta_\mu=i_\mu^*\eta$, we have
$$\flat(Y)[v]=d\eta_\mu(Y(p),[v])+\eta_\mu(Y(p))\eta_\mu([v])=d\eta(X_H(x),v)+\eta(X_H(x))\eta(v).$$
but $X_H$ is the Hamiltonian vector field associated to $H$, so
$$d\eta(X_H(x),v)+\eta(X_H(x))\eta(v)=\flat(X_H)(v)=d_xH(v)-(\RR(H)(x)+H(x))\eta(v).$$
If we prove that 
\begin{equation}\label{ladeR}
\RR(H)(x)=\RR_\mu(H_\mu)(p),
\end{equation}
and
\begin{equation}\label{ladeH}
d_xH(v)=d_pH_\mu([v]),
\end{equation}
we are done, because trivially $H(x)=H_\mu(p)$ and $\eta(v)=\eta_\mu([v])$ and so \eqref{relacionYHmu} is satisfied. For \eqref{ladeR} we use \eqref{H2},\eqref{relacionreebs} and the chain rule, so
\begin{align*}
\RR(H)(x)&=d_xH(\RR(x))=d_x(H_\mu\circ\pi_\mu)(\RR(x))\\&=d_pH_\mu(T_x\pi_\mu(\RR(x)))=d_pH_\mu(\RR_\mu(p))=\RR_\mu(H_\mu)(p).
\end{align*}
Similarly,
$$d_xH(v)=d_x(H_\mu\circ\pi_\mu)(v)=d_pH_\mu(T_x\pi_\mu(v))=d_pH_\mu([v]),$$
so we get what we wanted and the proof is finished.
\end{proof}

\label{sec2.3}
\chapter{Symplectification and its relation with reduction}\label{chapter3}
\section{Symplectification of contact manifolds}
During the last sixty years, symplectic manifolds have been thoroughly studied. This is not the case of contact manifolds, a much more recent field of study. In this context, symplectification \cite{ibanez1997co} of contact manifolds is an extremely useful tool, as it permits to study contact geometry problems with well-known symplectic techniques.  

Let $M$ be a $(2n+1)$-dimensional manifold and $\eta$ a $1$-form on $M$. Consider the product manifold $M\times \R$, with projection $p:M\times \R\rightarrow M$, and the $1$-form $\alpha=-e^t\eta$. Note that we should write $p^*\eta$ but we use just $\eta$ to not complicate the notation unnecessarily. Then, 
\begin{equation}
\Omega=-d\alpha=e^t(d\eta+dt\wedge\eta),
\end{equation}
defines a closed $2$-form on $M\times \R$. We have the following.

\begin{prop} $(M,\eta)$ is a contact manifold if and only if $(M\times\R,\Omega)$ is a symplectic manifold.
\end{prop}
\begin{proof} Since $\eta$ is a $1$-form, $\eta\wedge\eta=0$, thus
\begin{equation}\label{Omega}
\Omega^{n+1}=e^{(n+1)t}(d\eta^{n+1}+(n+1)dt\wedge\eta\wedge d\eta^{n})=(n+1)e^{(n+1)t}dt\wedge\eta\wedge d\eta^{n}.
\end{equation}
Note that $d\eta^{n+1}$ is a $(2n+2)$-form in a $(2n+1)$-dimensional manifold, and hence zero.

But the closed $2$-form $\Omega$ is symplectic if and only if $\Omega^{n+1}$ is not zero, and by \eqref{Omega}, this is true if and only if $\eta\wedge d\eta^{n}\neq 0$, that is, if and only if $\eta$ is a contact form.
\end{proof}

Next, we study the relation between Legendrian submanifolds of $M$ and Lagrangian submanifolds of $M\times \R$.

\begin{prop}\label{legendrianlagrangian} Let $(M,\eta)$ be a contact manifold, $N$ a submanifold of $M$, and $(M\times\R,\Omega)$ the symplectification of $M$. Then, $N$ is a Legendrian submanifold of $M$ if and only if $N\times \R$ is a Lagrangian submanifold of $M\times \R$.

\end{prop} 

\begin{proof}
Suppose that $N$ is a Legendrian submanifold. Then $\dim N=n$, and $\dim N\times\R=n+1=(1/2)\dim N\times\R$. Hence, by proposition \ref{nsymplectic}, it suffices to prove that $N\times\R$ is isotropic.

Let $(v,r), (w,s)\in T_xN\times\R=T_{(x,t)}\, N\times\R$. Since $N$ is Legendrian, proposition \ref{conflicto} gives $TN=TN^{\perp_\Lambda}=\ker (\eta_0)\cap \ker (d\eta_0)$, so $\eta(v)=\eta(w)=d\eta(v,w)=0$. Then we have
\begin{align*}
\nonumber \Omega((v,r),(w,s))&=e^t\left[d\eta((v,r),(w,s))+dt\wedge\eta((v,r),(w,s))\right]\\
\nonumber &=e^t\left[d\eta(v,w)+dt(r)\eta(w)-dt(s)\eta(v)\right]\\
&=e^t\left[d\eta(v,w)+r\eta(w)-s\eta(v)\right]=0.
\end{align*}
Thus $T_{(x,t)}\,N\times\R \subset (T_{(x,t)}\,N\times\R)^{\perp}$ and $N\times\R$ is isotropic.

Now assume that $N\times\R$ is Lagrangian. Then $\dim N\times\R=n+1$, and $\dim N=n$. Using proposition \ref{ncontact}, we are reduced to showing that $N$ is isotropic. Again by \ref{ncontact}, it suffices to prove that $N$ is horizontal.

Let $v\in T_xN, s\in \R$. Then $(v,0),(0,s)\in T_{(x,t)}\, N\times\R$ and beacuse $N\times\R$ is Lagrangian,
\begin{align*}
0= \Omega((v,0),(0,s))&=e^t\left[d\eta((v,0),(0,s))+dt\wedge\eta((v,0),(0,s))\right]\\
\nonumber &=e^t\left[d\eta(v,0)+dt(0)\eta(0)-dt(s)\eta(v)\right]\\
&=-e^tdt(s)\eta(v),
\end{align*}
Since this is true for all $s\in \R$, we deduce that $\eta(v)=0$ and then $N$ is horizontal.
\end{proof}

The following proposition shows that, given a coisotropic submanifold in $M$, one can obtain a coisotropic submanifold in $M\times\R$. In addition, the corresponding characteristic distributions are strongly related.

\begin{prop}\label{leaves} Let $(M,\eta)$ be a contact manifold, $N$ a coisotropic submanifold of $M$ which consists of vertical points, and $(M\times\R,\Omega)$ the symplectification of $M$. Then, $N\times \R$ is a coisotropic submanifold of $M\times\R$. Furthermore, if $x\in N$, $t\in \R$ and $N$ does not have horizontal points, we have
\begin{equation}\label{perp}
(T_{(x,t)}\,N\times\R)^{\perp}=(T_xN)^{\perp_\Lambda}\times\lbrace 0\rbrace.
\end{equation} 

\end{prop}

\begin{proof} First, we will prove that
\begin{equation}\label{contenido}
(T_xN)^{\perp_\Lambda}\times\lbrace 0\rbrace\subset (T_{(x,t)}\,N\times\R)^{\perp}. 
\end{equation}

Let $v\in (T_xN)^{\perp_\Lambda}$. We will show that $(v,0)\in (T_{(x,t)}\,N\times\R)^{\perp} $. By \ref{conflicto}, $\eta(v)=0$ and $d\eta(v,w)=0$ for all $w\in T_xN$. Let $(w,s)\in T_xN\times\R=T_{(x,t)}\, N\times\R$, then

\begin{align*}
\nonumber \Omega((v,0),(w,s))&=e^t\left[d\eta((v,0),(w,s))+dt\wedge\eta((v,0),(w,s))\right]\\
\nonumber &=e^t\left[d\eta(v,w)+dt(0)\eta(w)-dt(s)\eta(v)\right]\\
&=e^t\left[d\eta(v,w)-s\eta(v)\right]=0,
\end{align*}
and hence $(v,0)\in (T_{(x,t)}\,N\times\R)^{\perp}$.

Now suppose that $N$ is horizontal. By proposition \ref{ncontact}, $N$ is also isotropic, and hence Legendrian. Then, using proposition \eqref{legendrianlagrangian}, we get that $N\times\R$ is Lagrangian, thus coisotropic, so we are done.

Consider now the case when $N$ is not horizontal. We will prove \eqref{perp} arguing with dimensions. By remark  \ref{dimensioncomplement}, we get
$$\dim T_xN^{\perp_\Lambda}=2n+1-\dim T_xN,$$
but $(M\times\R,\Omega)$ is a $(2n+2)$-dimensional symplectic manifold, thus
$$\dim\,(T_{(x,t)}\,N\times\R)^{\perp}=2n+2-\dim\,(T_xN\times\R).$$
Hence,
$$\dim\, (T_{(x,t)}N\times\R)^{\perp}=\dim T_xN^{\perp_\Lambda}=\dim\,(T_xN)^{\perp_\Lambda}\times\lbrace 0\rbrace.$$
This, together with \eqref{contenido}, gives \eqref{perp}.

Finally, $(T_{(x,t)}\,N\times\R)^{\perp}=(T_xN)^{\perp_\Lambda}\times\lbrace 0\rbrace\subset T_{(x,t)}\,N\times\R$, so $N\times\R$ is coisotropic.
\end{proof}

\begin{obs}\label{V=U*t}
Equation \eqref{perp} has a stricking consequence: the leaves of the characteristic distribution of $N\times \R$ coincide with those of $N$, at each level $N\times \lbrace t\rbrace$ of $N\times\R$. More precisely, if $V$ is a leaf of the foliation given in $N\times \R$, then there exists $U$ leaf of the foliation in $N$ and $t\in \R$ such that $V=U\times\lbrace t\rbrace$.  $\hfill\diamondsuit$
\end{obs}

\section{Commutativity of symplectification and coisotropic reduction}\label{sec3.2}
Now we turn to the main problem of the present work. Let $(M,\eta)$ be a contact manifold, $N$ a coisotropic submanifold of $M$ which consists of vertical points, and $(M\times\R,\Omega)$ the symplectification of $M$. By \ref{leaves},  $N\times \R$ is a coisotropic submanifold of $M\times\R$. We need to determine whether the coisotropic reduction of $(M\times\R,\Omega)$ by $N\times\R$ coincides with the symplectification of the coistropically reduced space $(N/TN^{\perp_\Lambda},\tilde{\eta})$. To do so, we will construct a symplectomorphism between the two spaces. The following diagram describes the situation. 
$$
\xymatrix {
(M,\eta);\, N\subset M \ar[rrrr]^{ \text{symplectification}} \ar[ddd]_{\text{coisotropic\, reduction}} &&&& (M\times \R,\Omega);\, N\times\R\subset M\times\R \ar[dd]^{\text{coisotropic\, reduction}}\\ \\ &&&&  (N\times\R/T(N\times\R)^\perp, \tilde{\Omega}) \ar@{=}[d]^{\text{\, symplectomorphism?}} \\ (N/TN^{\perp_\Lambda},\tilde{\eta}) \ar[rrrr]^{\text{symplectification}} &&&& (N/TN^{\perp_\Lambda}\times \R,\bar{\Omega}),}
$$
where $\tilde{\Omega}$ is the symplectic form given by the coisotropic reduction of $(M\times\R,\Omega)$, and $\bar{\Omega}$ is the symplectic form obtained after the symplectification of the contact manifold $(N/TN^{\perp_\Lambda},\tilde{\eta})$. 
Let $p\in N\times\R/T(N\times\R)^\perp$ and let $(x,t)\in N\times \R$ such that $p=\pi_\mu(x,t)$, where $\pi:N\times \R\rightarrow N\times\R/T(N\times\R)^\perp$ is the canonical projection. We write $p=[x,t]$. Let $\tau:N\rightarrow N/TN^{\perp_\Lambda}$ be the other canonical projection, and write $\tau(x)=[x]_\Lambda$. We define,
\begin{align}
\nonumber H:N\times\R/T(N\times\R)^\perp &\rightarrow N/TN^{\perp_\Lambda}\times \R, \\
p \quad\quad &\mapsto \quad ([x]_\Lambda,t).
\end{align}
The diagram below clarifies the situation.
%$$\xymatrix {x\in N \ar@{|->}[ddd] && (x,t)\in N\times\R \ar@{<-|}[ll]\\ \\ &&  [x,t]\in N\times\R/T(N\times\R)^\perp \ar@{|->}[uu] \ar@{|->}[d]^{H} \\ [x]_\Lambda\in N/TN^{\perp_\Lambda} \ar@{|->}[rr] &&([x]_\Lambda,t)\in N/TN^{\perp_\Lambda}\times \R .}$$
$$
\xymatrix {
x\ar@{|->}[dd] \ar@{<-|}[rr] && (x,t)  \\ && [x,t] \ar@{|->}[u] \ar@{|->}[d]^{H} \\ [x]_\Lambda \ar@{|->}[rr] && ([x]_\Lambda,t).}
$$

\begin{teo}\label{ELTEOREMA1} The map $H$ just defined is a symplectomorphism, thus  $(N\times\R/T(N\times\R)^\perp, \tilde{\Omega})$ and $(N/TN^{\perp_\Lambda}\times \R,\bar{\Omega})$ are symplectomorphic symplectic spaces.
\end{teo}
\begin{proof}
First we must show that $H$ is well-defined. Let $(y,s)\in  N\times\R$ such that $[y,s]=[x,t]=p$. By remark \ref{V=U*t}, $s=t$ and both $x$ and $y$ lie on the same leaf of the foliation of $N$ given by $TN^{\perp_\Lambda}$, thus $[x]_\Lambda=[y]_\Lambda$. Hence, $([y]_\Lambda,s)=([x]_\Lambda,t)$, so $H$ is well-defined.

Similarly, it can be verified that $H$ is injective. In fact, if $H(p)=H(p')$ for $p=[x,t],p'=[y,s]\in N\times\R/T(N\times\R)^\perp$, then $([x]_\Lambda,t)=([y]_\Lambda,s)$, so $s=t$ and $x,y$ belong to the same leaf of the foliation of $N$. Again by remark \ref{V=U*t}, we obtain that $[x,t]=[y,s]$, thus $H$ is injective. Since $H$ is trivially surjective, we conclude that it is a bijection.  
\\

Next we prove that $H$ is differentiable. We do it locally. Let $k=\dim T_xN$ and $r=\dim T_xN^{\perp_\Lambda}$. Clearly $r\leq k$ because $N$ is coisotropic. Let $x_0\in N$ and take $(x^1,\ldots,x^k)$ local coordinates on $N$ around $x_0$. We can choose these coordinates to be adapted to the foliation given by $TN^{\perp_\Lambda}$, that is, such that the leaves of the foliation are defined locally by
$$x^{r+1}=\text{constant},\ldots,\, x^k=\text{constant}.$$
It is clear then that $(x^{r+1},\ldots,x^k)$ are local coordinates on $N/TN^{\perp_\Lambda}$ around $[x_0]_\Lambda$. Now let $t_0\in \R$ and define
\begin{align*}
\tilde{x}^i:N\times \R&\rightarrow \R,\\(x,t)&\mapsto x^i(x).
\end{align*}
Then $(\tilde{x}^1,\ldots,\tilde{x}^k,\tilde{t})$ are local coordinates on $N\times\R$ about $(x_0,t_0)$. Moreover, by \ref{V=U*t}, these are adapated coordinates to the foliation given by $T(N\times \R)^\perp$, that is,
$$\tilde{x}^{r+1}=\text{constant},\ldots,\, \tilde{x}^k=\text{constant},\,\tilde{t}=\text{constant},$$
are local equations for the leaves. Hence, $(\tilde{x}^{r+1},\ldots,\tilde{x}^{k},\tilde{t})$ are local coordinates of the leaf space $N\times\R/T(N\times\R)^\perp$ around $[x_0,t_0]$. Notice that similarly, $(\bar{x}^{r+1},\ldots,\bar{x}^k,\bar{t})$ defined by
\begin{align*}
\bar{x}^i:N/TN^{\perp_\Lambda}\times \R&\rightarrow \R,\\([x]_\Lambda,t)&\mapsto x^i(x),
\end{align*}
are coordinates on $N/TN^{\perp_\Lambda}\times \R$ about $([x_0]_\Lambda,t_0)=H([x_0,t_0])$. Written in these coordinates, $H$ is just the identity map, thus $H$ is differentiable. 
\\

Finally, we show that $H$ is a symplectomorphism, that is, $H^*\bar{\Omega}=\tilde{\Omega}$. Let $p\in N\times\R/T(N\times\R)^\perp$ and let $(x,t)\in N\times \R$ such that $p=[x,t]$. Take $u,v\in T_p (N\times\R/T(N\times\R)^\perp)$. We must show that $H^*\bar{\Omega}(p)(u,v)=\tilde{\Omega}(p)(u,v)$, or equivalently,

$$\bar{\Omega}(H(p))(d_pH(u),d_pH(v))=\tilde{\Omega}(p)(u,v).$$

Since $\pi$ is a submersion, there exist $\tilde{u},\tilde{v}\in T_xN$ and $t_1,t_2\in \R$ such that $T_{(x,t)}\pi(\tilde{u},t_1)=u, T_{(x,t)}\pi(\tilde{v},t_2)=v$. Applying theorem \ref{symcoired} to $\Omega,\tilde{\Omega}$ gives $\pi^*\tilde{\Omega}=\Omega$, and then
\begin{align} \label{symplectomorphism1}
\nonumber \tilde{\Omega}(p)(u,v)&=\tilde{\Omega}(\pi(x,t))(T_{(x,t)}\pi(\tilde{u},t_1),T_{(x,t)}\pi(\tilde{v},t_2))\\
\nonumber &=\pi^*\tilde{\Omega}(x,t)((\tilde{u},t_1),(\tilde{v},t_2))=\Omega(x,t)((\tilde{u},t_1),(\tilde{v},t_2))\\&=e^t\left[d\eta(x)(\tilde{u},\tilde{v})+t_1\eta(x)(\tilde{v})-t_2\eta(x)(\tilde{u})\right].
\end{align}

Using now \ref{concoired}, one has $\tau^*\tilde{\eta}=\eta$, thus
$$\eta(x)(\tilde{u})=\tau^*\tilde{\eta}(x)(\tilde{u})=\tilde{\eta}([x]_\Lambda)(T_x\tau(\tilde{u})),$$
and the same remains true for $\tilde{v}$. Since $\tau^*d\tilde{\eta}=d\tau^*\tilde{\eta}=d\eta$, we also have,
$$d\eta(x)(\tilde{u},\tilde{v})=\tau^*d\tilde{\eta}(x)(\tilde{u},\tilde{v})=d\tilde{\eta}([x]_\Lambda)(T_x\tau(\tilde{u}),T_x\tau(\tilde{v})).$$

Introducing the last two equalities in \eqref{symplectomorphism1}, we get
\begin{align*}
\tilde{\Omega}(p)(u,v)&=e^t[d\tilde{\eta}([x]_\Lambda)(T_x\tau(\tilde{u}),T_x\tau(\tilde{v}))+t_1\tilde{\eta}([x]_\Lambda)(T_x\tau(\tilde{v}))-t_2\tilde{\eta}([x]_\Lambda)(T_x\tau(\tilde{u}))]\\ &= e^t[d\tilde{\eta}([x]_\Lambda)(T_x\tau(\tilde{u}),T_x\tau(\tilde{v}))+dt\wedge\tilde{\eta}([x]_\Lambda,t)((T_x\tau(\tilde{u}),t_1),(T_x\tau(\tilde{v}),t_2))]\\&=\bar{\Omega}([x]_\Lambda,t)((T_x\tau(\tilde{u}),t_1),(T_x\tau(\tilde{v}),t_2)).
\end{align*}
But by definition we have $H\circ\pi(x,t)=([x]_\Lambda,t)=(\tau(x),t)$, and by the chain rule
$$(T_x\tau(\tilde{u}),t_1)=d_{(x,t)}H\circ\pi(\tilde{u},t_1)=d_{\pi(x,t)}(T_{(x,t)}\pi(\tilde{u},t_1))=d_pH(u).$$
Finally, 
$$\tilde{\Omega}(p)(u,v)=\bar{\Omega}([x]_\Lambda,t)((T_x\tau(\tilde{u}),t_1),(T_x\tau(\tilde{v}),t_2))=\bar{\Omega}(H(p))(d_pH(u),d_pH(v)),$$
thus $H^*\bar{\Omega}=\tilde{\Omega}$ and $H$ is a symplectomorphism.

We conclude that $(N\times\R/T(N\times\R)^\perp, \tilde{\Omega})$ and $(N/TN^{\perp_\Lambda},\tilde{\eta})$ are symplectomorphic symplectic spaces, as we claimed.
\end{proof}

\label{sec3.1}
\section{Commutativity of symplectification and reduction}
In this section we present a result similar to theorem \ref{ELTEOREMA1}, but now concernig reduction via momentum mappings. Let $(M,\eta)$ be a contact manifold, $\Phi:G\times M\rightarrow M$ a contact action of a given Lie group $G$ on $M$, and $J:M\rightarrow \mathfrak{g}^*$ the asociated momentum map given by definition \ref{contact momentum def}. Consider now the symplectification $(M\times \R,\Omega)$ of $M$. We define the symplectificated action $\tilde{\Phi}$ by,  
\begin{align*}
\tilde{\Phi}:G\times (M\times\R)&\rightarrow M\times\R,\\ (g,(x,t))&\mapsto (\Phi_g(x),t).
\end{align*}
To guarantee that this action has an asociated momentum mapping it suffices to prove that it is in the assumptions of proposition \ref{exact symplectic momentum map}. 

\begin{prop} The symplectificated action $\tilde{\Phi}$ is symplectic. Furthermore, it leaves $\alpha=-e^t\eta$ invariant and so $\tilde{J}:M\times \R$ defined by
\begin{equation}\label{tildejota}
\tilde{J}(x,t)\xi=\alpha(\xi_{M\times \R}(x,t)),
\end{equation}
is an Ad$^*$ equivariant momentum map for the symplectificated action.
\end{prop}
\begin{proof} Let $g\in G$. We will show that $\tilde{\Phi}_g^*\alpha=\alpha$. Take $(x,t)\in M\times \R$ and $(u,s)\in T_xM\times \R=T_{(x,t)}(M\times\R)$. We have,
\begin{align*}
\tilde{\Phi}_g^*\alpha (x,t)(u,s)&=\alpha(\tilde{\Phi}_g(x,t))(d_{(x,t)}\tilde{\Phi}_g(u,s))=\alpha(\Phi_g(x),t)(d_x\Phi_g(u),s) \\ &=-e^t\eta(\Phi_g(x))(d_x\Phi_g(u))=-e^t\eta(x)(u)=\alpha(x)(u,s).
\end{align*}
where we have used that $\Phi_g^*\eta=\eta$ beacuse $\Phi$ is a contact action. 

Now, it is clear that
$$\Omega=-d\alpha=-d\tilde{\Phi}_g^*\alpha=\tilde{\Phi}_g^*(-d\alpha)=\tilde{\Phi}_g^*\Omega,$$
thus the symplectificated action is symplectic and preserves $\alpha$. The last statement is then a direct consequence of proposition \ref{exact symplectic momentum map}.
\end{proof}

Both momentum maps $J$ and $\tilde{J}$ are related by a simple expression.
\begin{lema} Let $(x,t)\in M\times\R$ and $\xi\in \mathfrak{g}$. Then we have,
\begin{equation}\label{lemamalo}
\tilde{J}(x,t)(\xi)=e^tJ(x)(\xi).
\end{equation}
\end{lema}
\begin{proof} By definition of $\xi_{M\times\R}$ and $\tilde{\Phi}$, we have
\begin{align*}
\xi_{M\times \R}(x,t)&=\left.\dfrac{\partial}{\partial r}\right|_{r=0} (\tilde{\Phi}_{exp(r\xi)}(x,t))=\left.\dfrac{\partial}{\partial r}\right|_{r=0}(\Phi_{exp(r\xi)}x,t)\\&=\left( \left.\dfrac{\partial}{\partial r}\right|_{r=0} (\Phi_{exp(r\xi)}x)\, ,\,0 \right)=(\xi_M(x),0).
\end{align*}
Then, using definitions \eqref{tildejota} and \eqref{contact momentum formula},
$$\tilde{J}(x,t)(\xi)=\alpha(x,t)(\xi_{M\times \R}(x,t))=\alpha(x,t)(\xi_M(x),0)=-e^t\eta(\xi_M(x))=e^tJ(x)(\xi).$$
\end{proof}

Now, take $\mu\in \mathfrak{g}^*$ a regular value of both $J$ and $\tilde{J}$, and suppose that it is a fixed point of $J$ under the coadjoint action, that is, $G_\mu=G$. Assume also that the action of $G$ is free and proper on $J^{-1}(\mu)$ and $\tilde{J}^{-1}(\mu)$. Then $M_\mu=J^{-1}(\mu)/G$ and $\tilde{M}_\mu=\tilde{J}^{-1}(\mu)/G$ are well-defined contact and symplectic manifolds, respectively. The natural question that arises is whether the symplectification of the contact manifold $M_\mu$ coincides with the symplectic manifold $\tilde{M}_\mu$. The following diagram depics de situacion.
$$
\xymatrix {
J^{-1}(\mu)\subset M \ar[rrrr]^{ \text{symplectification}} \ar[ddd]_{\text{contact\, reduction}} &&&& \tilde{J}^{-1}(\mu)\subset M\times\R \ar[dd]^{\text{symplectic\, reduction}}\\ \\ &&&&  \tilde{M}_\mu=\tilde{J}^{-1}(\mu)/G \ar@{=}[d]^{\text{\, symplectomorphism?}} \\ M_\mu=J^{-1}(\mu)/G \ar[rrrr]^{\text{symplectification}} &&&& M_\mu\times \R,}
$$
where we have omitted the contact and symplectic forms because they are the standard ones for this construction. Notice that by  lemmas \ref{lemahojaorbitacontacto} and \ref{lemahojaorbita}, iii), we have that $J^{-1}(\mu)$ and $\tilde{J}^{-1}(\mu)$ are coisotropic submanifolds. Moreover, the contact and symplectic reductions coincide with coisotropic reduction, by direct use of formulas \eqref{hojaorbitacontacto} and \eqref{hojaorbita}. If we proved that $J^{-1}(\mu)\times\R=\tilde{J}^{-1}(\mu)$, we would be exactly in the situation described in section \ref{sec3.2}. Unfortunately this is not true in general, as a consequence of \eqref{lemamalo}. But for $\mu=0$, equation \eqref{lemamalo} implies that $J^{-1}(0)\times\R=\tilde{J}^{-1}(0)$. We have proved the following theorem.

\begin{teo}
The symplectification of $M_0$ and the reduced space $\tilde{M}_0$ are symplectomorphic symplectic spaces.
\end{teo}

What happens in the general case cannot be deduced by coisotropic reduction argumentes so it is matter of future research.

\label{sec3.3}
\bibliography{bibliografia}
\bibliographystyle{plain}

\end{document}